\documentclass[twoside]{IEEEtran} 

\IEEEoverridecommandlockouts                              
\usepackage{amssymb,amsfonts,mathptmx,stmaryrd}
\usepackage[pdftex]{graphicx}
\pdfoutput=1
\usepackage[cmex10]{amsmath}
\usepackage{array}

\newcommand{\eps}{\ensuremath{\varepsilon}}

\newcommand{\ome}{\ensuremath{\omega}}
\newcommand{\Ome}{\ensuremath{\Omega}}
\newcommand{\vro}{\ensuremath{\varrho}}

\newcommand{\eea}{\end{eqnarray}}
\newcommand{\bea}{\begin{eqnarray}}
\newcommand{\nn}{\ensuremath{\nonumber}} 

\newcommand{\bean}{\begin{eqnarray*}} 
\newcommand{\eean}{\end{eqnarray*}} 

\newcommand{\eeq}{\end{equation}}
\newcommand{\beq}{\begin{equation}}

\newcommand{\mrm}{\ensuremath{\mathrm}}

\newcommand{\llb}{\ensuremath{\llbracket}} 
\newcommand{\rrb}{\ensuremath{\rrbracket}} 
\newcommand{\lp}{\ensuremath{\left (}} 
\newcommand{\rp}{\ensuremath{\right )}} 
\newcommand{\conn}{\ensuremath{\leftrightsquigarrow}}

\newcommand{\WW}{\ensuremath{\mathbb{Z}_{\geq 0}}}
\newcommand{\NN}{\ensuremath{\mathbb{N}}}
\newcommand{\RR}{\ensuremath{\mathbb{R}}}

\newcommand{\mbA}{\ensuremath{\mathbf{A}}}
\newcommand{\mbB}{\ensuremath{\mathbf{B}}}
\newcommand{\mbD}{\ensuremath{\mathbf{D}}}
\newcommand{\mbX}{\ensuremath{\mathbf{X}}}
\newcommand{\mbY}{\ensuremath{\mathbf{Y}}}
\newcommand{\mbQ}{\ensuremath{\mathbf{Q}}}

\newcommand{\mbS}{\ensuremath{\mathbf{S}}}

\newcommand{\mbT}{\ensuremath{\mathbf{T}}}
\newcommand{\mbP}{\ensuremath{\mathbf{P}}}
\newcommand{\mbK}{\ensuremath{\mathbf{K}}}
\newcommand{\mbC}{\ensuremath{\mathbf{C}}}

\newcommand{\mbF}{\ensuremath{\mathbf{F}}}
\newcommand{\mbO}{\ensuremath{\mathbf{O}}}
\newcommand{\mbI}{\ensuremath{\mathbf{I}}}
\newcommand{\mbR}{\ensuremath{\mathbf{R}}}
\newcommand{\mbH}{\ensuremath{\mathbf{H}}}

\newcommand{\mcH}{\ensuremath{\mathcal{H}}}

\newcommand{\mcC}{\ensuremath{\mathcal{C}}}
 
\newcommand{\mcF}{\ensuremath{\mathcal{F}}}

\newcommand{\mcP}{\ensuremath{\mathcal{P}}}
\newcommand{\mcT}{\ensuremath{\mathcal{T}}}
\newcommand{\mcO}{\ensuremath{\mathcal{O}}}
\newcommand{\mcQ}{\ensuremath{\mathcal{Q}}}
\newcommand{\mcX}{\ensuremath{\mathcal{X}}}

\newcommand{\mrh}{\ensuremath{\mathrm{h}}} 
\newcommand{\mrf}{\ensuremath{\mathrm{f}}} 

\newcommand{\dia}{\ensuremath{\mathrm{diam}}}
\newcommand{\mrH}{\ensuremath{\mathrm{H}}} 
\newcommand{\mrI}{\ensuremath{\mathrm{I}}} 
\newcommand{\mrIs}{\ensuremath{\mathrm{I}_*}}

\newcommand{\ess}{\ensuremath{\mathrm{ess}}}
\newcommand{\mrT}{\ensuremath{\mathrm{T}}}

\newcommand{\Xh}{\ensuremath{\hat{X}}}

\newtheorem{dfn}{Definition}[section] 
\newtheorem{thm}{Theorem}[section] 
\newtheorem{lem}{Lemma}[section] 
 
\newtheorem{cor}{Corollary}[section]

\title{\LARGE \bf
A Nonstochastic Information Theory\\
for Communication and State Estimation
}

\author{Girish N. Nair
\thanks{Published in {\em IEEE Trans. Automatic Control}, vol. 58, no. 6, pp. 1497--1510, 2013. This work was supported by Australian Research Council grant DP110102401.
A  preliminary version appeared in \cite{nairICCA11}.  
G.N. Nair is with the Department of Electrical and Electronic Engineering, 
        University of Melbourne, VIC 3010, Australia,  
	tel: +61-3-8344-6701, fax: +61-3-8344-6678, email:  
        {\tt\small gnair@unimelb.edu.au}}%
}

\begin{document}

\thispagestyle{empty}

\maketitle

\begin{abstract}

In communications, unknown variables are usually modelled as random variables, 
and concepts such as independence, entropy and information are defined in terms of the underlying probability distributions. 
In contrast, control theory often treats uncertainties and disturbances as bounded unknowns having no statistical structure. 
The area of networked control combines both fields,  raising the question of whether it is possible to 
construct meaningful analogues of stochastic concepts such as independence, Markovness,  entropy and information 
without assuming a probability space. This paper introduces a framework for doing so, 
leading  to the construction of a  {\em maximin information} functional for nonstochastic variables. 
It is shown that the largest maximin information rate through a memoryless, error-prone channel 
in this framework coincides with the block-coding zero-error capacity of the channel.
Maximin information is then used to derive  tight conditions for uniformly estimating the 
 state of a linear time-invariant system  over such a  channel,
paralleling recent results of Matveev and Savkin.
\end{abstract}


\section{INTRODUCTION}
\label{introsec}

This paper has two motivations.
The first arises out of the analysis of networked control systems \cite{antsaklisSpecial04}, which 
combine the two different disciplines of communications and control.
In communications systems, 
 unknown quantities are usually modelled as random variables  
(rv's),  and
central concepts such as independence, Markovness, entropy and Shannon information are defined stochastically. 
One reason for this is that they 
are generally prone to 
electronic circuit noise, which obeys physical laws  yielding well-defined distributions. 
In addition, communication systems are often used many times,
and in everyday applications each phone call and data byte 
may not be important.
Consequently, the system designer need only ensure good performance in an average or expected sense
 - e.g. small  bit error rates and large signal-to-noise average power ratios. 

In contrast, control is often used in safety- or mission-critical applications
where  performance must be guaranteed every time a plant is used, not just on average.
Furthermore, in plants that contain mechanical and chemical
components, the dominant disturbances may not necessarily arise from circuit noise,
and may not follow a well-defined  probability distribution.
Consequently, control theory often treats uncertainties and
disturbances as bounded unknowns or signals without statistical structure.
Networked control thus raises natural questions  
of whether it is possible to construct useful analogues of the stochastic concepts mentioned above,
{\em without assuming a probability space}.

Such questions are not new and some answers are available.
For instance, if an rv has known range but unknown distribution,
then its uncertainty may be quantified by the logarithm of  
the cardinality or  Lebesgue measure of this range.
This leads to the notions of {\em Hartley entropy} $\mrH_0$ \cite{hartley} for discrete variables 
and  {\em R\'{e}nyi differential 0th-order entropy} $\mrh_0$ \cite{renyiBerkeley} 
for continuous variables.  
A related construction is the $\eps$-entropy, 
which is the log-cardinality of  the smallest partition of
a given metric space such that each partition set 
has diameter no greater than $\eps>0$ \cite{kolmogorov59,jagerman69,donoho00}.
None of these concepts require any statistical structure.

Using these notions,  nonstochastic measures of information
can be constructed.
For instance, in \cite{shinginAUTO12} 
the difference between the marginal and worst-case conditional R\'{e}nyi entropies was 
taken to define  a  nonstochastic, asymmetric information functional,
and  used to study  feedback control over errorless digital channels. 
In  \cite{klirBook},
{\em information transmission} was defined    
symmetrically as the difference between the
sum of the marginal and the joint Hartley entropies of a pair of discrete variables.
Continuous variables with convex ranges admitted a similar construction,
but with $\mrH_0$   replaced by a projection-based, isometry-invariant functional.
Although both these  definitions  possess many natural properties,
their  wider operational relevance
is unclear.   
This contrasts with Shannon's theory, which is 
intimately connected to quantities of practical significance in engineering, 
such as the minimum  and maximum bit-rates for reliable compression and transmission \cite{shannon}.

The second, seemingly unrelated motivation comes from the study of {\em zero-error capacity} $C_0$ \cite{shannonTIT56,kornerTIT98}
in communications. The zero-error capacity of a stochastic discrete channel is the largest block-coding 
rate possible across it that ensures zero probability of decoding error. 
This is a more stringent concept than the {\em (ordinary) capacity} $C$ \cite{shannon},
defined to be the  highest block-coding rate such that the probability of a decoding error is arbitrarily small.
The famous {\em channel coding theorem} \cite{shannon} 
states that the capacity of a stochastic, memoryless channel
 coincides with the highest rate of Shannon information across it, 
a purely  intrinsic quantity. 
In \cite{wolfITW04}, an analogous identity for $C_0$ was found in terms of  
the Shannon entropy of the `largest' rv common to the channel input and output.
However, it is known that $C_0$ does not depend on the values
of the non-zero transition probabilities in the channel and can be defined without any reference to a probabilistic framework.
This strongly suggests that 
$C_0$  should be expressible as
the maximum rate of a suitably defined  {\em nonstochastic} information index.

This paper has four main contributions. In section  \ref{forsec}, a formal framework for modelling nonstochastic {\em uncertain variables (uv's)} is proposed,
leading to analogues of probabilistic ideas such as independence and  Markov chains. 
In section  \ref{infsec}, the  concept of {\em maximin information} $\mrIs$ is introduced to quantify  how much the uncertainty in one 
uv can be reduced by observing another. Two characterizations of $\mrIs$ are given here, and shown to be equivalent. 
In section  \ref{channelsec}, the notion of an error-prone, stationary memoryless channel is defined within the uv framework,
and it is proved in Theorem  \ref{mainthm} that the zero-error capacity $C_0$ of any such channel
coincides with the largest possible rate of maximin information across it.  
Finally, it is shown in section  \ref{stateestsec}
how $\mrIs$ can be used to find a tight condition (Theorem  \ref{capthm}) that describes
whether or not  the state of a noiseless linear time-invariant (LTI) system can be estimated with specified exponential 
uniform accuracy over an erroneous channel.  
A tight criterion for the achievability of uniformly  bounded estimation
errors is also derived for when uniformly bounded additive disturbances are present (Theorem  \ref{capthm2});
a similar result was derived in \cite{matveevIJC07}, using probability arguments 
but no information theory. In a nonstochastic setting, maximin information thus serves to  
delineate the limits of reliable communication and LTI state estimation 
over error-prone channels.

\section{Uncertain Variables}
\label{forsec}
The key idea in the framework proposed here  
is to keep the probabilistic convention of 
regarding an unknown variable as a mapping $X$ from some underlying {\em sample space} $\Ome$ to
a set $\mbX$ of interest. 
For instance,  in a dynamic system each {\em sample} $\ome\in\Ome$ may  be identified with  a particular combination of initial states
and exogenous noise signals, and gives rise to a realization $X(\ome)$ denoted by lower-case $x\in\mbX$.
Such a mapping $X$ is called an {\em uncertain variable (uv)}. As in probability theory, the dependence on $\omega$ is usually suppressed for conciseness,
so that a statement such as $X\in\mbK$ means $X(\ome)\in\mbK$.  
However, unlike probability theory, the formulation presented here
assumes neither a family of measurable subsets of $\Ome$, nor a measure on them.

Given another uv $Y$ taking values in $\mbY$, write 
\bea
\llb X\rrb &:=&\{  X(\ome): \ome\in\Ome\}   ,\label{defset}\\
 \llb X|y\rrb   &:= & \left\{ X(\ome): Y(\ome)=y, \ome\in\Ome\right\}, \label{defcondset}\\ 
\llb X,Y\rrb    &:= & \left\{ \left (X(\ome),Y(\ome)\right ):\ome\in\Ome\right\}.  
\label{defjointset}
\eea
Call $\llb X\rrb $ the {\em marginal range} of $X$, 
$\llb X|y\rrb $ its  {\em conditional range} given (or {\em range conditional on}) $Y=y$, 
and $\llb X,Y\rrb   $, the {\em joint range} of $X$ and $Y$.  
With some abuse of notation, denote the family of conditional ranges (\ref{defcondset}) as  
\beq
\llb X|Y\rrb   := \left\{\llb  X|y\rrb   : y\in \llb Y\rrb \right\}, 
\label{defcover}
\eeq
with empty sets omitted. 
In the absence of stochastic structure, the  uncertainty associated with $X$ given all possible realizations  of 
$Y$ is described by the set-family $\llb  X|Y\rrb  $.
Notice that $\cup_{\mbB\in\llb  X|Y\rrb   }\mbB=\llb  X\rrb   $, i.e.~$\llb  X|Y\rrb   $ is an  $\llb  X\rrb   ${\em-cover}.    
In addition, 
\beq
\llb  X,Y\rrb    = 
\bigcup_{y\in\llb  Y\rrb } 
\llb  X|y\rrb   \times\{y\}, 
\label{XY}
\eeq
i.e. the joint range is fully determined by the conditional and marginal ranges in a manner
that parallels the relationship between joint, conditional and marginal probability distributions.

Using this basic framework, a nonstochastic analogue of  statistical independence 
can be defined:
\begin{dfn}[Unrelatedness] 
\label{unrelateddfn}
A collection of uncertain variables $Y_1,\ldots , Y_m$ is said to be {\em (unconditionally) unrelated}  
if 
\[
\llb  Y_1,\ldots , Y_m\rrb   =\llb  Y_1\rrb   \times\cdots\times\llb  Y_m\rrb.
\]
They are said to be {\em conditionally unrelated given} 
(or {\em unrelated conditional on}) $X$ 
if 
\[
\llb  Y_1,\ldots , Y_m|x\rrb   =\llb  Y_1|x\rrb   \times\cdots\times\llb  Y_m|x\rrb   ,  \ \  x\in\llb  X\rrb.
\]
\end{dfn}

$\diamondsuit$

Like independence, unrelatedness has an alternative characterization in terms of conditioning:
\begin{lem}  
\label{unrelatedlem}
Given uncertain variables $X,Y,Z$,  
\begin{enumerate}
\item 
$Y,Z$ are unrelated (Definition \ref{unrelateddfn}) iff 
the conditional range 
\[
\llb  Y|z\rrb   =\llb  Y\rrb   , \ \  z\in\llb  Z\rrb   .
\]
\item $Y,Z$ are unrelated conditional on $X$ iff 
\[
\llb  Y|z,x\rrb   =\llb  Y|x\rrb   , \ \  (z,x)\in\llb  Z,X\rrb.
\]
\end{enumerate}
\end{lem}

{\em Proof:} Trivial. $\Box$

\begin{figure}[!t]
\centering
\includegraphics[width=3.4in]{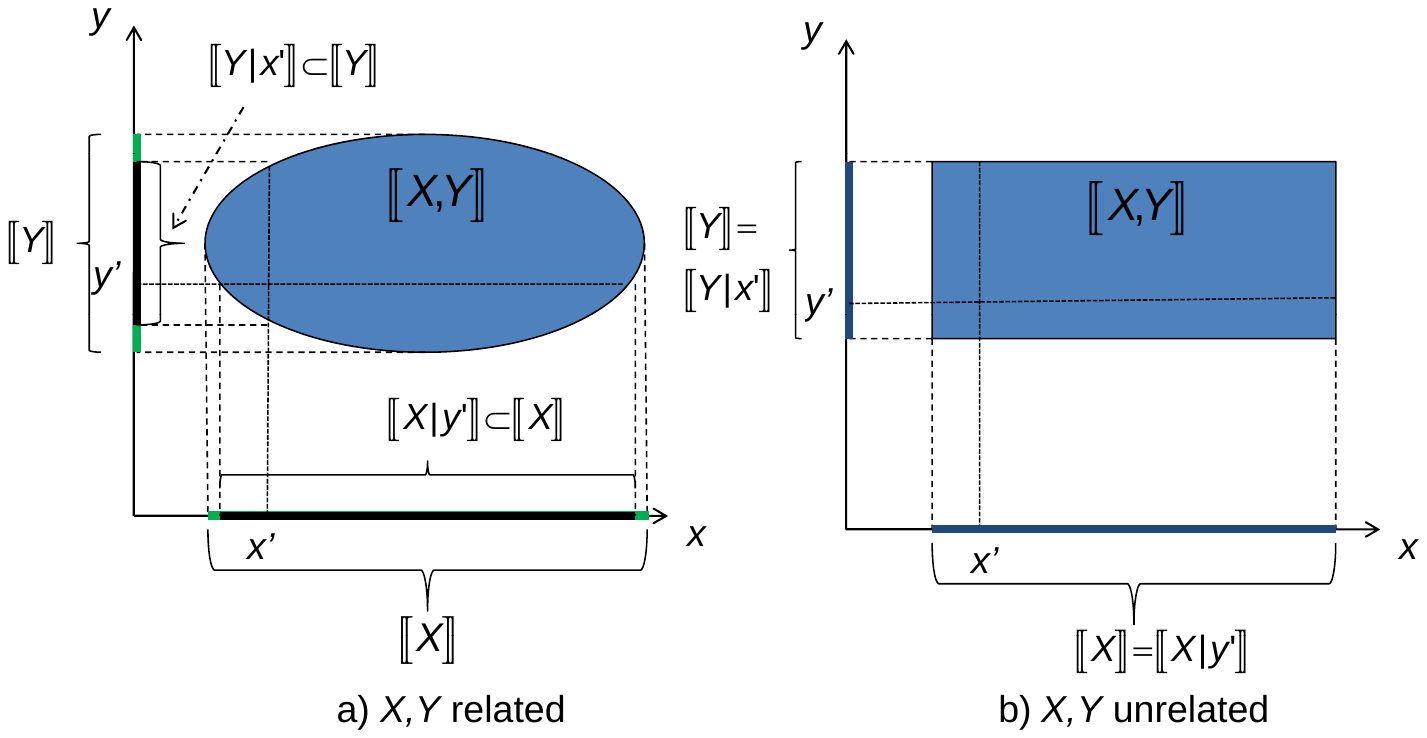}
\caption{Examples of joint and marginal ranges for related and unrelated uv's.}
\label{relatedfig} 
\end{figure}

{\bf Example:} Figure \ref{relatedfig}a) illustrates the case of two related uv's $X$ and $Y$.
Observe that the joint range $\llb X,Y\rrb$
is strictly contained in the Cartesian product $\llb X\rrb\times\llb Y\rrb$ of marginal ranges.
In  addition, for some values  $x'\in\llb X\rrb$ and $y'\in\llb Y\rrb$, the conditional ranges $\llb X|y'\rrb$ and $\llb Y|x'\rrb$
are strictly contained in the marginal ranges $\llb X\rrb$ and $\llb Y\rrb$, respectively.

In contrast, Figure \ref{relatedfig}b) depicts the ranges when $X$ and $Y$ are unrelated.
The joint range 
now coincides with $\llb X\rrb\times\llb Y\rrb$,
and $\llb X|y'\rrb$ and $\llb Y|x'\rrb$
coincide with $\llb X\rrb$ and $\llb Y\rrb$ respectively,
for every  $x'\in\llb X\rrb$ and $y'\in\llb Y\rrb$.

It is  to see that for any uv's $X,Y_1,\ldots , Y_m$,   
\beq
\llb  X|y_1,\ldots , y_m\rrb   \subseteq\llb  X|y_1\rrb   \cap\cdots\cap\llb  X|y_m\rrb   ,
\label{subsplit}
\eeq
for all $(y_i)_{i=1}^m\in\llb  Y_1\rrb   \times\cdots\llb  Y_m\rrb   $ and $i\in [1:m]$. 
Equality is  possible under extra hypotheses: 
\begin{lem}
\label{splitlem}
Let $X,Y_1,\ldots, Y_m$ be uncertain variables s.t.
$Y_1,\ldots , Y_m$ are unrelated conditional on $X$ (Definition \ref{unrelateddfn}).
Then  $\forall (y_1,\ldots , y_m)\in\llb  Y_1,\ldots , Y_m\rrb   $,
\beq
\llb  X|y_1,\ldots , y_m\rrb   =\llb  X|y_1\rrb   \cap\cdots\cap\llb  X|y_m\rrb.
\label{split}
\eeq
\end{lem}

{\em Proof:}  
See appendix  \ref{splitlempf}. $\Box$

The second item  in Lemma \ref{unrelatedlem} motivates the following definition:
\begin{dfn}[Markov Uncertainty Chains] 
\label{markovdfn}
The uncertain variables $X$, $Y$ and $Z$ are said to form a 
{\em Markov uncertainty chain} $X\leftrightarrow Y\leftrightarrow Z$ if $X,Z$ are unrelated conditional on $Y$
(Definition \ref{unrelateddfn}).
\end{dfn}

$\diamondsuit$

{\bf Remarks:} By the symmetry of Definition \ref{unrelateddfn}, 
$Z\leftrightarrow Y\leftrightarrow X$
is also a Markov uncertainty chain.

Before closing this section, it is noted that the framework developed above is not 
equivalent to treating input variables with known, bounded ranges 
as uniformly distributed rv's. 
Such an approach is still probabilistic, and the output rv's 
may  have nonuniform distributions despite the uniform inputs.
In contrast,  in the uv model here, only the ranges are considered, 
and no distributions are derived at any stage.

For instance, consider an additive bounded noise channel with output $Y=X+N$, 
where the input $X$ and noise $N$ range on the interval $[-0.5,0.5]$. 
If $X$ and $N$ are taken to be mutually independent, uniform rv's,
then $Y$ has a triangular distribution on $[-1,1]$,
with small values of $Y$  more probable than large ones.
 However, if  $X$ and $N$ are treated as unrelated uv's,
then all that can be inferred about $Y$ is that it has range $[-1,1]$,
with all values in this range being equally possible.

Naturally, this lack of statistical structure does not suit all applications.
 However, as discussed in section \ref{introsec},
such structure is often excess to requirements, e.g. in problems with worst-case objectives and bounded variables
as in section \ref{stateestsec}.
A uv-based approach is arguably more natural in these settings.

\section{Maximin Information}
\label{infsec}

The framework introduced above  
is now used to define a nonstochastic analogue $\mrIs$ of 
Shannon's mutual information functional.
Two  characterizations
of $\mrIs$ are developed  and shown to be equivalent (Definition \ref{Isdefn} and Corollary \ref{taxicabcor}). 

Throughout this section, $X,Y$ are arbitrary uncertain variables (uv's) 
with marginal ranges $\llb X\rrb$ and $\llb Y\rrb$
(\ref{defset}),
joint range $\llb X,Y\rrb$ (\ref{defjointset}), and 
conditional range family $\llb X|Y\rrb$   (\ref{defcover}).
Set cardinality is
denoted by  $|\cdot|$, with the value $\infty$  permitted, and all logarithms are to base 2.

\subsection{Previous Work}

It is useful to first recall 
the nonprobabilistic formulations of {\em entropy} and information mentioned in section  \ref{introsec}.
Though originally defined in different settings,
for the sake of notational coherence they are discussed here using the uv framework of section  \ref{forsec}.
 
In loose terms, the entropy of a variable quantifies 
the  prior uncertainty associated with it.
For discrete-valued  $X$, this uncertainty may be captured by  
the (marginal) {\em Hartley} or {0-entropy} 
\beq
\mrH_0[ X]   :=\log|\llb  X\rrb   |\in [0,\infty],  
\label{Hartley}
\eeq
If $\llb  X\rrb   $ has Lebesgue measure $\mu$ on $\RR^n$,
then the (marginal)  {\em R\'{e}nyi differential 0-entropy}  
 is defined as    
\beq
\mrh_0[  X]   :=\log\mu\llb  X\rrb   \in [-\infty , \infty]. 
\label{defh0}
\eeq
A related construction is the $\eps$-entropy,
which is the log-cardinality of the smallest partition of
a given metric space such that each partition set 
has diameter no greater than $\eps>0$ \cite{kolmogorov59}.
None of these concepts require a probability space.

Two distinct notions of information have been proposed based on the 0-entropies above.
In  \cite{shinginAUTO12}, a worst-case approach is taken
to first define the {\em  (conditional) 0-entropy of $X$ given} 
 $Y$ as  
\beq  
\mrH_0[   X|Y]   := \ess\sup_{y\in \llb  Y\rrb   } \log\left |\llb  X|y\rrb \right |   
\in [0 ,\infty].   
\label{cond0ent}
\eeq
If every set in the family $\llb  X|Y\rrb   $ is $\mu$-measurable on $\RR^n$,
then the {\em (conditional) differential 0-entropy of $X$ given} $Y$ is  
\beq  
\mrh_0[  X|Y]   := \ess\sup_{y\in \llb  Y\rrb   } \log\mu\llb  X|y\rrb    
\in [-\infty ,\infty].   
\label{cond0ent2}
\eeq
Noting that Shannon information can be expressed as 
the difference between the marginal and conditional entropies,
a nonstochastic {\em 0-information} functional $\mrI_0$ is   
then defined  in \cite{shinginAUTO12} as 
\beq
\mrI_0[X;Y]   := 
\mrH_0[  X]   -\mrH_0[ X|Y]  
 \equiv \ess\inf_{y\in\llb  Y\rrb } 
\log\left (\frac{|\llb  X\rrb   |}{|\llb X|y\rrb|}\right )
\label{0inf1}
\eeq
if $X$ is discrete-valued with $\mrH_0[X|Y]<\infty$, and 
as
\beq  
\mrI_0[X;Y]  :=
\mrh_0[  X]   -\mrh_0[  X|Y]    
\equiv \ess\inf_{y\in\llb  Y\rrb  } \log\left (\frac{\mu\llb  X\rrb   }{\mu\left (\llb X|y\rrb\right )}\right )
\label{0inf2}
\eeq 
 if $X$ is continuous-valued with $\mrh_0[X;Y]<\infty$.
In other words, the 0-information that can be gained about $X$ from
 $Y$ is the worst-case log-ratio of the prior to posterior uncertainty set sizes.\footnote{Note that in 1965, 
Kolmogorov had defined    
$\log\left |\llb  X|y\rrb \right |$  as a `combinatorial' conditional entropy and 
the log-ratio $\log\left (|\llb  X\rrb   |/|\llb X|y\rrb|\right )$ as a measure of information gain. 
However, these quantities have the defect of depending on the observed value $Y=y$,
 and thus are associated with a specific posterior uncertainty
set $\llb X|y\rrb$. In contrast, (\ref{cond0ent})--(\ref{0inf2}) and (\ref{Isimple}) are
functions of the {\em family} $\llb X|Y\rrb$ of all possible posterior uncertainty sets.}
     
The  definition above is inherently asymmetric, i.e. $\mrI_0[X;Y]\neq \mrI_0[Y;X]$.
A different and symmetric nonstochastic information index 
 had been previously proposed in \cite{klirBook}.   
In that formulation, 
a conditional entropy was first defined
as the difference between the
joint and marginal Hartley entropies, in analogy with Shannon's theory.
The {\em information transmission} $\mrT[X;Y]$ was 
then defined as the difference between the marginal and conditional entropies,
yielding the symmetric formula 
\[
\mrT[X;Y]:=\mrH_0[X]+ \mrH_0[Y] - \mrH_0[X,Y].
\]  
Continuous variables with convex ranges admitted a similar construction,
with $\mrH_0$   replaced not with $\mrh_0$ but a projection-based, 
isometry-invariant functional.

Though  these concepts are intuitively appealing and 
share some desirable properties with Shannon information,
they have two weaknesses.
Firstly, 
they do not treat 
continuous- and discrete-valued uv's in a unified way.
In particular, 
it is unclear how to apply the  approach of \cite{klirBook}
to mixed pairs of variables,
e.g. a digital symbol encoding a continuous state,
or to continuous variables with nonconvex ranges.

Secondly and more importantly, 
their operational relevance for problems involving communication
has not been generally established.
While the worst-case log-ratio approach of \cite{shinginAUTO12}
has been used to find 
minimum bit rates for stabilization over an errorless digital channel,
it is not obvious how to apply it  
if transmission errors occur. 

For these reasons, an alternative approach is pursued in the 
remainder of this section. 

\subsection{$\mrIs$ via the Overlap Partition}
\label{overlapsubsec}

The nonstochastic information index $\mrIs$  
proposed in this subsection quantifies the information 
that can be gained about $X$ from $Y$ in terms of 
certain structural properties  of the family $\llb X|Y\rrb$ of posterior uncertainty sets.
These properties are described below:

\begin{dfn}[Overlap Connectedness/Isolation] 
\label{overlapdefn}
\hfill 
\begin{enumerate}
\item A pair of points $x$ and $x'\in\llb X\rrb$ is called $\llb X|Y\rrb$-{\em overlap connected}, denoted $x\conn x'$,  
if $\exists$ a finite sequence $\{\llb X|y_i\rrb\}_{i=1}^{n}$ of conditional ranges  
such that $x\in\llb X|y_1\rrb$, $x'\in\llb X|y_n\rrb$ and 
each conditional range has nonempty intersection with its predecessor, i.e.  
$\llb X|y_i\rrb\cap\llb X|y_{i-1}\rrb\neq\emptyset$, for each $i\in[2,\ldots , n]$.

\item A set $\mbA\subseteq\llb X\rrb$ is called $\llb X|Y\rrb$-overlap connected  if 
every pair of points in $\mbA$ is overlap connected.

\item A pair of sets $\mbA,\mbB$ is called $\llb X|Y\rrb$-{\em overlap isolated} if 
no point in $\mbA$ is overlap connected with any point in $\mbB$.

\item An  $\llb X|Y\rrb$-{\em overlap isolated partition} (of $\llb X\rrb$) 
is a partition of $\llb X\rrb$ where every pair of  distinct member-sets is overlap isolated.

\item An $\llb X|Y\rrb$-{\em overlap partition} 
is an overlap-isolated partition each member-set of which is overlap connected. 
\end{enumerate}
\end{dfn}
$\diamondsuit$


{\bf Remarks:}  
For conciseness, the qualifier $\llb X|Y\rrb$- will often dropped when 
there is no risk of confusion about the conditional range family of interest.   
Note that any point or set is automatically overlap connected with itself. 
In addition, $x'$ lies in the same overlap partition set as $x$ iff $x'\conn x$.

Symmetry and transitivity
guarantee that a unique overlap partition always exists: 
\begin{lem}[Unique Overlap Partition] 
\label{overlaplem}
There is a unique $\llb X|Y\rrb$-overlap partition 
of  $\llb X\rrb$  (Definition \ref{overlapdefn}),  denoted $\llb X|Y\rrb_*$. 
Every set $\mbC\in\llb X|Y\rrb_*$ is expressible as
\beq
\mbC=\{x\in\llb X\rrb: x\conn\mbC\}=\bigcup_{\mbB\in\llb X|Y\rrb:\mbB\conn\mbC}\mbB.
\label{overlapsets}
\eeq

Furthermore, every $\llb X|Y\rrb$-overlap isolated partition $\mcP$ of  $\llb X\rrb$ 
satisfies
\beq
|\mcP|\leq |\llb X|Y\rrb_*|,
\label{maxcard}
\eeq
with equality iff $\mcP= \llb X|Y\rrb_*$.
\end{lem}  

{\em Proof:}
See appendix  \ref{overlaplempf}.
$\Box$

{\bf Remarks:} The self-referential identities in (\ref{overlapsets}) are 
needed to prove certain key results later.
The first equality says that each element $\mbC$ of the overlap partition
coincides with the set of all points that are overlap connected with it.   
The second states that every such $\mbC$ is expressible as a union of elements of 
the set family $\llb X|Y\rrb $.

Observe that from Definition \ref{overlapdefn}, overlap-isolated partitions  
are precisely those partitions $\mcP$ of $\llb X\rrb$ with the property that
 every conditional range $\llb X|y\rrb$ lies {\em entirely} inside  one 
member set $\mbP\in\mcP$. In other words, 
each possible observation $y\in\llb Y\rrb$ unambiguously identifies exactly one 
partition set $\mbP$ containing $x$. Equivalently, these partition sets can be thought of
as defining a discrete-valued function, or quantizer, on $\llb X\rrb$.  The more sets there are in $\mcP$,
the more distinct values this quantizer can take, and so
the more refined the knowledge that can be unequivocally gained about $X$.

By the result above, $\llb X|Y\rrb_*$ is precisely the overlap-isolated partition of maximum cardinality.
This leads naturally to the definition below: 
  

\begin{dfn}
\label{Isdefn}
The {\em maximin information} between $X$ and $Y$ is defined as    
\beq
\mrIs[X;Y]:= \log\left |\llb  X|Y\rrb_* \right |,
\label{Isimple}
\eeq
where 
$\llb X|Y\rrb_*$ is the unique $\llb X|Y\rrb$-overlap 
partition of $\llb X\rrb$ (Lemma  \ref{overlaplem}). 
\end{dfn}

$\diamondsuit$

{\bf Remarks:} By the discussion above, $\mrIs[X;Y]$  
represents  the most refined knowledge that can be gained about $X$ 
from observations of $Y$.    
Note that this definition applies to both continuous- and discrete-valued uv's.
Also note that the self-information $\mrIs[X;X]$ is identical to $\mrH_0[X]$.

\begin{figure}[!t]
\centering
\includegraphics[width=3.4in]{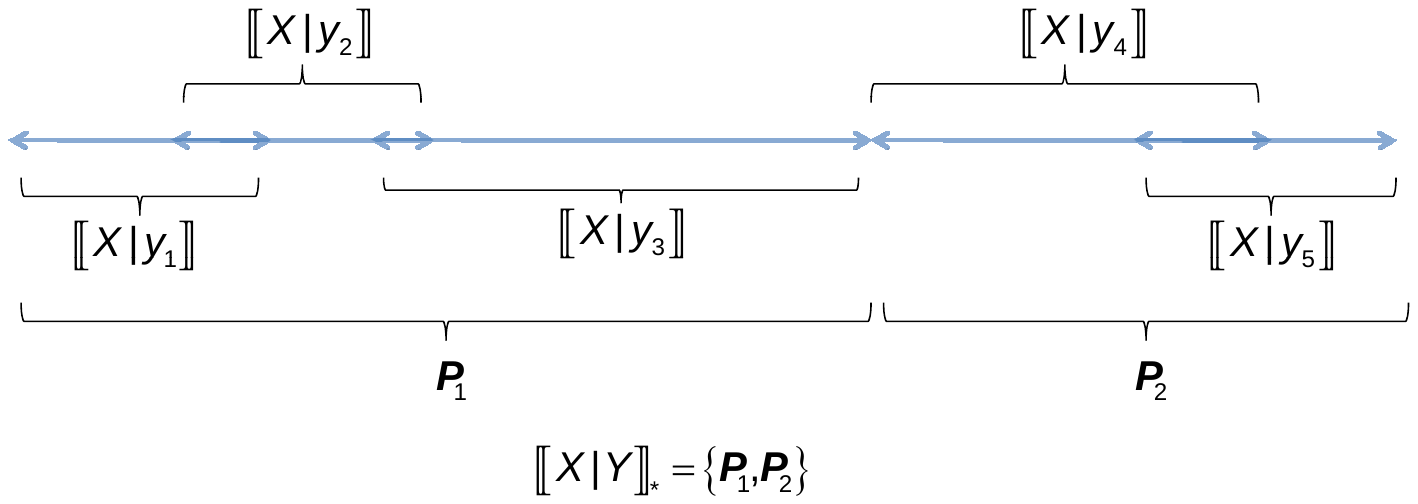}
\caption{Example of  an overlap partition.}
\label{overlapfig} 
\end{figure}

{\bf Example:}
Consider uv's $X$ and $Y$ with the one-dimensional conditional range family
$\llb X|Y\rrb=\left\{\llb X|y_i\rrb: i=1,\ldots ,5\right\}$
and overlap partition $\llb X|Y\rrb_*=
\{\mbP_1,\mbP_2\}$  
depicted in Figure \ref{overlapfig}.
Observe that any pair  of points in $\mbP_1$ or in $\mbP_2$ is overlap connected,
and no point in $\mbP_1$ is overlap connected to a point in $\mbP_2$.  
Also note that  $\{\mbP_1,\mbP_2\}$ is the finest partition of $\llb X\rrb$
having member sets that can always be unambiguously determined from $Y$;
a partition with  larger cardinality would necessarily  
contain two or more neighbouring partition sets  intersected by the same
posterior set $\llb X|y_i\rrb$, and the observation $Y=y_i$  
would then correspond to either partition set.
Thus the maximin information between $X$ and $Y$ is $\log |\llb X|Y\rrb_*|=\log 2=1$ bit. 

It is easy to  verify that $\mrIs\neq\mrI_0$.

{\bf Example:} Let $X$ and $Z$ be unrelated uv's with  $\llb  X\rrb   =\{  0,1\}   $
and $\llb  Z\rrb   =\{0,1\}$, and define the uv $Y$ by  
$Y=X$ if $Z=0$ and $Y=2$ if $Z=1$.
The family $\llb  Y|X\rrb$ consists of the sets 
$\llb Y|0\rrb = \{0,2\}$
and $\llb Y|1\rrb = \{1,2\}$.
The overlap partition $\llb  Y|X\rrb_*$ has only one set, $\{0,1,2\}$,
 so $\mrIs[Y;X]=\log 1 =0$.
However $\mrI_0[Y;X]=\log\frac{3}{2}$, since the largest cardinality of sets in $\llb  Y|X\rrb$ is 2.   

Finally, note that 
$\mrIs[X;Y]$ was originally defined  in  \cite{nairICCA11}
as 
\[
\sup_{\mbF\in\mcF\llb  X\rrb }\min_{\mbC\in\llb  X|Y\rrb_*  }\log\left (\frac{|\mbF|}{|\mbF\cap\mbC|}\right ),
\]
where  $\mcF\llb  X\rrb   $ is the family of all finite subsets of $\llb  X \rrb$;
hence the name `maximin' information.
This log-ratio characterization is close in spirit to (\ref{0inf1})--(\ref{0inf2})
and can be shown to be equivalent to (\ref{Isimple}). 
However, since it does not have as simple an interpretation as (\ref{Isimple}) 
and is not needed for any of the results here,
there will be no further discussion of it in what follows.   

\subsection{$\mrIs$ via the Taxicab Partition}
\label{taxicabsubsec}

The definition  of maximin information above is based purely on the conditional range family $\llb X|Y\rrb$.
As  $\llb Y|X\rrb$ will not generally be the same, it may seem that $\mrIs$ could be asymmetric in its arguments.  
However, it turns out that $\mrIs$ can be reformulated symmetrically in terms of the joint range $\llb X,Y\rrb$.
A few additional concepts are needed
in order to present this  characterization.

\begin{dfn}[Taxicab Connectedness/Isolation] 
\label{taxidefn}
\hfill 
\begin{enumerate}
\item A pair of points $(x,y)$ and $(x',y')\in\llb X,Y\rrb$ is called {\em taxicab connected}
if there is a {\em taxicab sequence} connecting them, i.e. a 
finite sequence $\{(x_i,y_i)\}_{i=1}^{n}$ of points in $\llb X,Y\rrb$ such that
 $(x_1,y_1) =(x,y)$, $(x_n,y_n)=(x',y')$ and 
  each point differs in at most one coordinate from its predecessor, 
i.e.  $y_i= y_{i-1}$ and/or $x_i = x_{i-1}$,  for each $ i\in [2,\ldots , n]$. 

\item A set $\mbA\subseteq\llb X,Y\rrb$ is called taxicab connected if 
every pair of points in $\mbA$ is taxicab connected in $ \llb X,Y\rrb$.

\item A pair of sets $\mbA,\mbB$ is called {\em taxicab isolated} if 
no point in $\mbA$ is taxicab connected in $ \llb X,Y\rrb$ to any point in $\mbB$.

\item A {\em taxicab-isolated partition (of $\llb X,Y\rrb$)}     
is a cover of $\llb X,Y\rrb$ such that 
 every pair of distinct sets in the cover is taxicab isolated. 

\item A {\em taxicab partition (of $\llb X,Y\rrb$)}     
is a taxicab-isolated partition of $\llb X,Y\rrb$  each member-set of which is taxicab connected. 
\end{enumerate}
\end{dfn}

$\diamondsuit$

\begin{figure}[!t]
\centering
\includegraphics[width=3.4in]{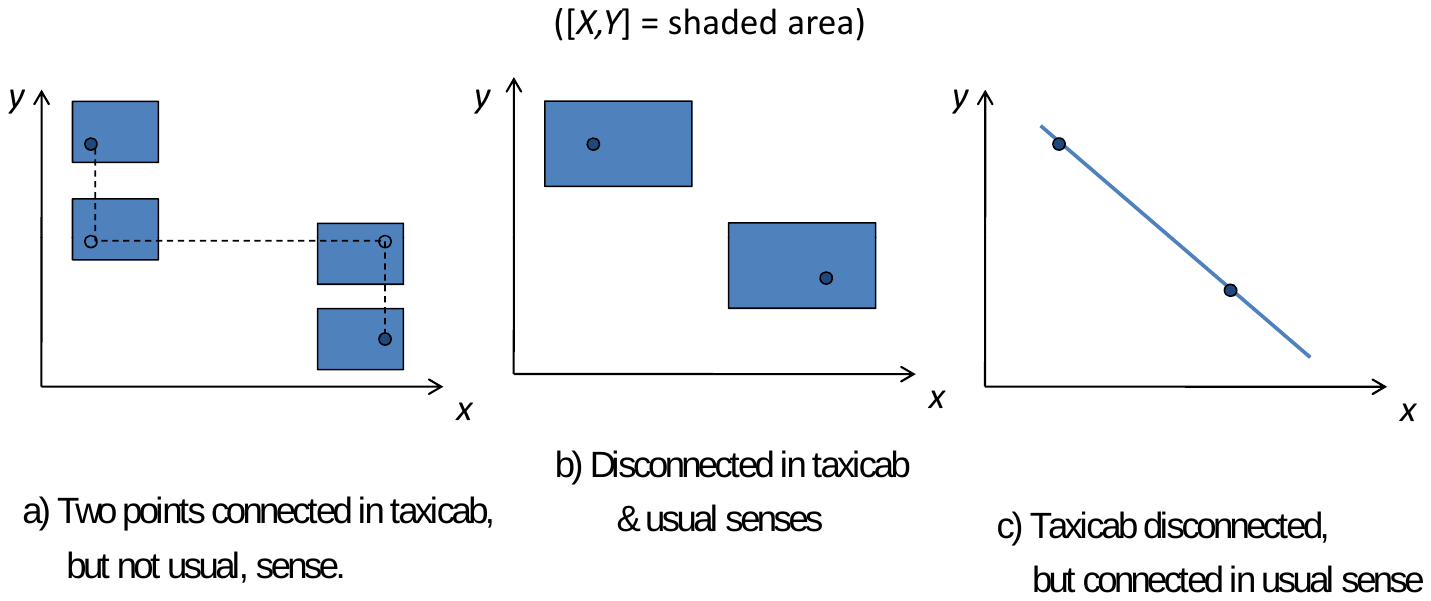}
\caption{Path- vs. taxicab-connectedness}
\label{taxicabfig} 
\end{figure}

{\bf Remarks:}
Note that any point or set is automatically taxicab connected with itself.
In addition, taxicab connectedness/isolation  in $\llb X,Y\rrb$  is identical to that in $\llb Y,X\rrb$, with 
the order of elements in each pair reversed.
Consequently, any taxicab-isolated partition of  $\llb X,Y\rrb$  is in one-to-one correspondence with one of $\llb Y,X\rrb$.

Taxicab-isolated partitions have the property that the
 particular member set $\mbT$ that contains   
a given  point $(x,y)$ is uniquely determined by $x$ and by $y$ alone.
The argument is by contradiction: if $x$ is associated with two sets $\mbT,\mbT'$ in the overlap-isolated
partition, i.e. $(x,y)\in\mbT$ and $(x,y')\in\mbT'$
for distinct $y,y'\in\llb Y\rrb$, then $\mbT$ and $\mbT'$ would be taxicab-connected
by the sequence $\lp (x,y),(x,y')\rp$.
In other words, the sets of a taxicab-isolated partition represent posterior knowledge 
that can always be agreed on by two agents 
who separately observe realizations of $X$ and $Y$.

\begin{lem}[Taxicab- $\Leftrightarrow$ Overlap-Connectedness] 
\label{connectlem}
Any two points $(x,y),(x',y')\in\llb X,Y\rrb$ are taxicab connected
(Definition \ref{taxidefn}) 
iff $x\conn x'$ in $\llb X|Y\rrb$ (Definition \ref{overlapdefn}). 

Thus any set $\mbA\subseteq\llb X,Y\rrb$ is taxicab connected 
iff its $x$-axis projection $\mbA^+\subseteq\llb X\rrb$ is overlap connected. 

Similarly, any two sets $\mbA,\mbB\subseteq\llb X,Y\rrb$ are taxicab isolated 
(Definition \ref{taxidefn}) 
iff $\mbA^+,\mbB^+\subseteq\llb X\rrb$ are overlap isolated
(Definition \ref{overlapdefn}). 
\end{lem}

{\em Proof:}
See appendix  \ref{connectlempf}. $\Box$

Due to this equivalence between the two notions of connectedness, the same symbol $\conn$ is used. 
The result below makes another link:
\begin{thm}[Unique Taxicab Partition] 
\label{partitionthm}
There  is  
a unique taxicab partition  (Definition \ref{taxidefn}) $\mcT[X;Y]$  of $\llb X,Y\rrb$
(\ref{defjointset}).

In addition, every taxicab-isolated partition $\mcQ$ of   $\llb X,Y\rrb$
satisfies
\beq
|\mcQ|\leq |\mcT[ X;Y]|,
\label{maxcardtaxi}
\eeq
with equality iff $\mcQ= \mcT[ X;Y]$.

Furthermore, a  one-to-one correspondence from
$\mcT[X;Y]$ to the  overlap partition $\llb X|Y\rrb_*$ (Lemma  \ref{overlaplem})
is obtained by projecting the sets of the 
former
 onto $\llb X\rrb$.
\end{thm}  

{\em Proof:}
See appendix  \ref{partitionthmpf}. $\Box$

The last statement of this theorem leads immediately to the following alternative characterization of maximin information: 
\begin{cor}[$\mrIs$ via Taxicab Partition] 
\label{taxicabcor}
The maximin information $\mrIs$ (\ref{Isimple}) satisfies the identity 
\[
\mrIs[X;Y] = \log\left |\mcT[X;Y]\right |,
\]
where  $\mcT[X;Y]$ is the unique taxicab partition  
 of $\llb X,Y\rrb$
(Theorem  \ref{partitionthm}).

Thus $\mrIs[X;Y]=\mrIs[Y;X]$. 
\end{cor}

$\diamondsuit$

{\bf Remarks:} From the discussion following Definition \ref{taxidefn}, the bound (\ref{maxcardtaxi})
means that $ \mcT[ X;Y]$ represents the {\em finest} 
 posterior knowledge that can be agreed on from individually observing $X$ and $Y$. 
The log-cardinality of  this partition has considerable intuitive appeal as an index  
of  information.
Indeed, if $X$ and $Y$ are discrete rv's,  then the elements of the taxicab partition
correspond to the {\em connected components} of the bipartite graph that describes  $(x,y)$ pairs with nonzero joint probability.
In \cite{wolfITW04}, the Shannon entropy of these connected components was called {\em zero-error information} 
and used to derive an intrinsic but stochastic characterization of the zero-error capacity $C_0$ of discrete memoryless channels.
Maximin information corresponds rather to the Hartley entropy of these connected components.  
In section  \ref{channelsec},  it  
will be seen to yield  an analogous {\em nonstochastic} characterization
that is valid for discrete- or continuous-valued channels. 

\begin{figure}[!t]
\centering
\includegraphics[width=3.4in]{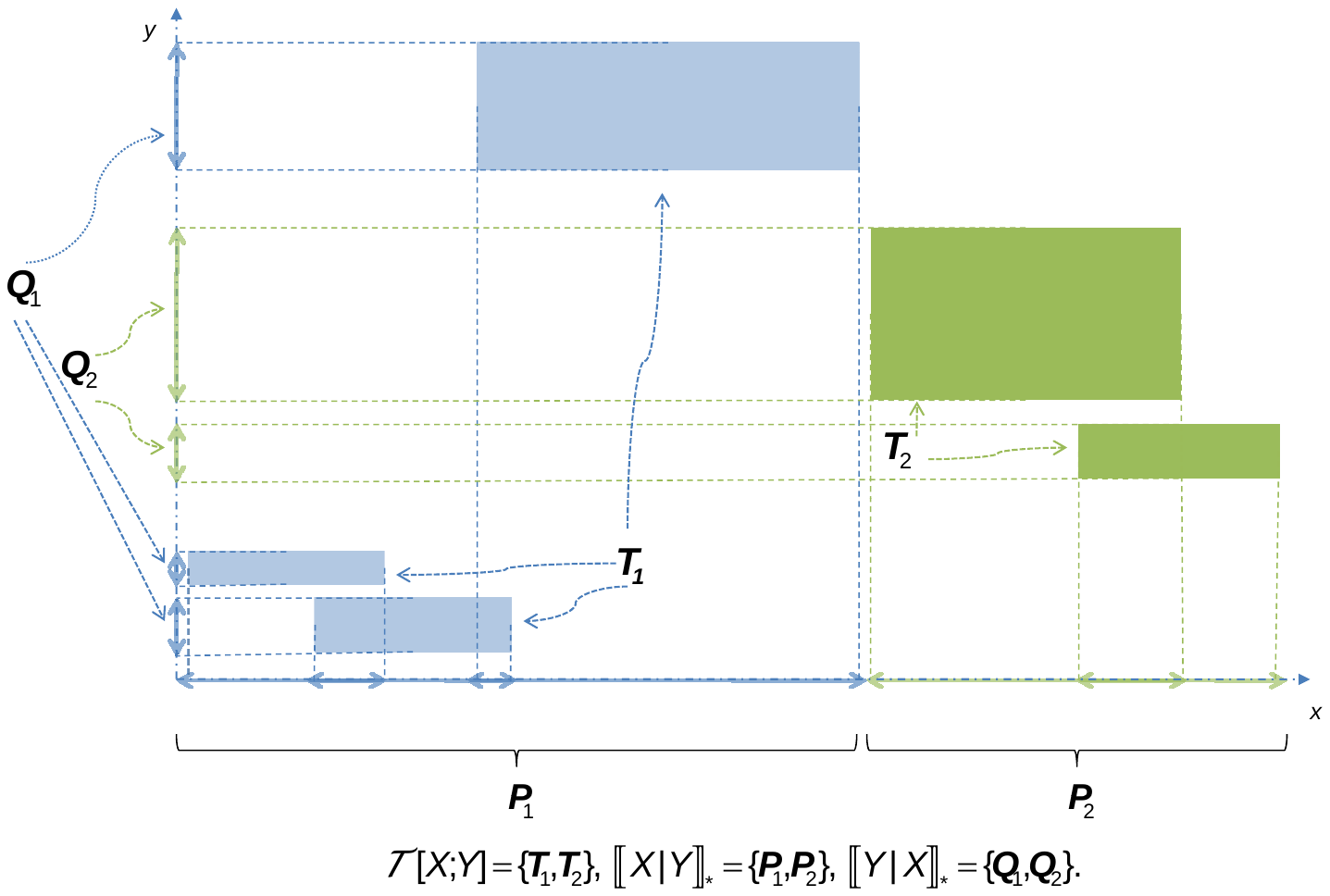}
\caption{Taxicab and Overlap Partitions}
\label{extrafig} 
\end{figure}

{\bf Example:}
The shaded regions in Figure \ref{extrafig} depict the joint range $\llb X,Y\rrb$ of uv's $X,Y$ 
having the conditional range  family $\llb X|Y\rrb$  
of Figure \ref{overlapfig}.
The taxicab partition $\mcT[X;Y]$ consists of the sets $\mbT_1$ and $\mbT_2$;
 it can be seen that every pair of points in each set is taxicab connected,
and no point in one set is taxicab-connected with a point in the other.
 Projecting $\mbT_1$ and $\mbT_2$ onto $\llb X\rrb$ yields
$\mbP_1$ and $\mbP_2$, the sets comprising the overlap partition
$\llb X|Y\rrb_*$. Similarly,   $\llb Y|X\rrb_*$ consists of the projections $\mbQ_1$ and $\mbQ_2$ of
$\mbT_1$ and $\mbT_2$ onto $\llb Y\rrb$.

If two agents observe $X$ and $Y$ separately,
then they will  always be able to agree on the index $Z\in\{1,2\}$
of the unique taxicab partition set $\mbT_Z$ that contains $(X,Y)$,
since it is also the index of the  overlap partition sets $\mbP_Z$ and $\mbQ_Z$ that contain $X$ and $Y$ respectively.
The amount of information they share is then  
$\log |\mcT[X;Y]| = \log |\llb X|Y\rrb_*| =\log |\llb Y|X\rrb_*|=1$ bit.

\subsection{Properties of Maximin Information}  
\label{propertysec}

Two important properties of maximin information are now established. 
These properties are also exhibited by Shannon information and 
will be needed to prove Theorem   \ref{capthm}. 

\begin{lem}[More Data Can't Hurt] 
\label{morelem}
The maximin information $\mrIs$ (\ref{Isimple}) satisfies
\beq
\mrIs[X;Y]    \leq  \mrIs[X;Y,Z].
\label{more}
\eeq
\end{lem}

{\em Proof:}
By Definition \ref{overlapdefn}, every set $\mbC\in\llb X|Y,Z\rrb_*$  
 is  overlap connected in $\llb X|Y,Z\rrb$.
As $\llb X|y,z\rrb\subseteq\llb X|y\rrb$, 
$\mbC$ is also overlap connected in $\llb X|Y\rrb$.
Pick a set  $\mbC'\in\llb X|Y\rrb_*$ that intersects $\mbC$.
As $\mbC'$ is overlap connected in $\llb X|Y\rrb$,
it  also $\conn\mbC$. Thus $\mbC\subseteq\mbC'$, since
by (\ref{overlapsets}) 
$\mbC'$ must include all points $\conn\mbC'$.
Consequently, there is only one $\mbC'$ for each $\mbC$.

Furthermore, since   $\llb X|Y,Z\rrb_*$
covers $\llb X\rrb$, every set of   $\llb X|Y\rrb_*$
must intersect and thus include some of its set(s).   
Thus the map $\mbC\mapsto\mbC'$ is a surjection
from $\llb X|Y,Z\rrb_*\to \llb X|Y\rrb_*$,
implying that  $|\llb X|Y,Z\rrb_*|\geq | \llb X|Y\rrb_*|$.
$\Box$

\begin{lem}[Data Processing] 
\label{DPlem}
If $X\leftrightarrow Y\leftrightarrow Z$ 
is a Markov uncertainty-chain (Definition \ref{markovdfn}),
then the maximin information $\mrIs$ (\ref{Isimple}) satisfies
\beq
\mrIs[ X;Z]    \leq  \mrIs[  X;Y].
\label{dp1}
\eeq
\end{lem}

{\em Proof:}
By Lemma \ref{morelem},
\[
\mrIs[ X;Z]    \leq  \mrIs[  X;Y,Z]
\stackrel{(\ref{Isimple})}{=}\log|\llb X|Y,Z\rrb|_*.
\]
By Definition \ref{markovdfn}, $\llb X|y,z\rrb=\llb X|y\rrb$
for every $\forall y\in\llb Y\rrb$ and $z\in\llb Z|y\rrb$,
so $\llb X|Y,Z\rrb_*=\llb X|Y\rrb_*$.
Substituting this into the RHS of the equation above and applying (\ref{Isimple}) again
completes the proof.
$\Box$

{\bf Remark:} By the symmetry of Markov uncertainty chains and maximin information,
 $\mrIs[ X;Z]    \leq  \mrIs[Y;Z]$.  

\subsection{Discussion}
\label{remarksec}

Maximin information is a more conservative index than Shannon information $I$.
For instance,  
Corollary \ref{taxicabcor}  implies that unrelated uv's must share 0 maximin information,
but the converse does not hold, unlike the analogous case with Shannon
information. This is because $\mrIs[X;Y]$  is the largest cardinality of $\llb X,Y\rrb$-partitions  
such that the unique partition set containing any realization $(x,y)$ can be determined by observing
either  $x$ or $y$ alone.  Even if $X$ and $Y$ are related, there may be no way to split the joint range into two or more  
sets 
   that are each unambiguously identifiable in this way.  

\begin{figure}[!t]
\centering
\includegraphics[width=3.4in]{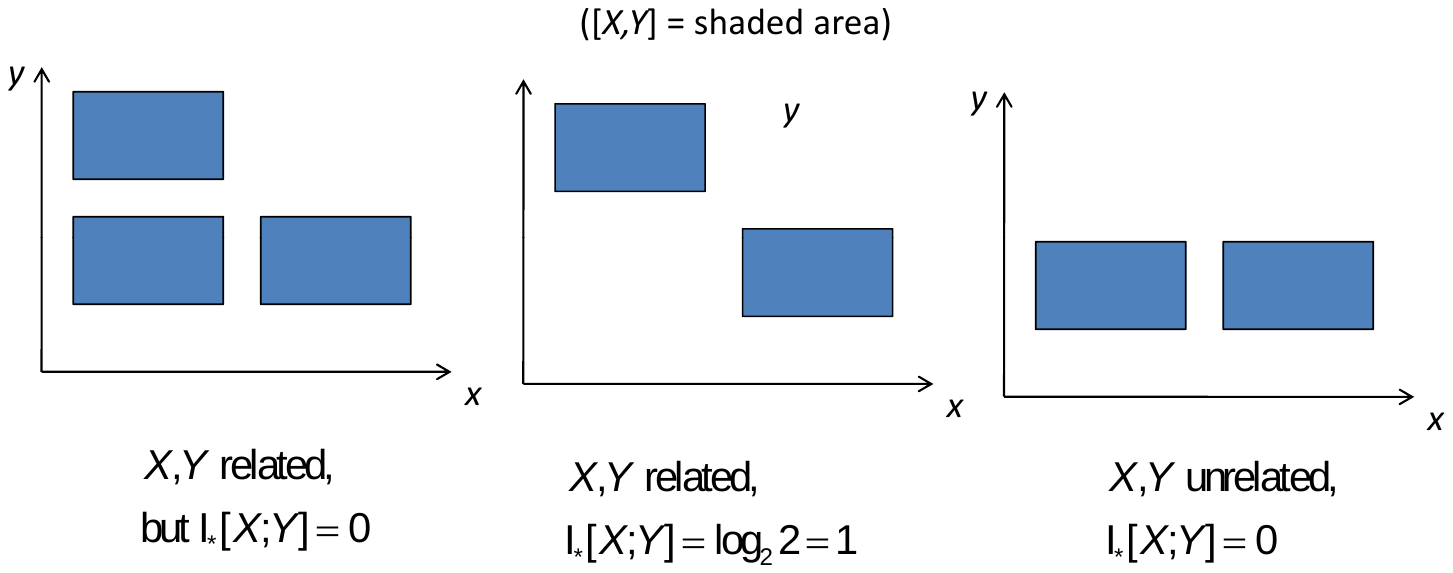}
\caption{Zero $\mrIs$ does not imply unrelatedness}
\label{zeroinffig} 
\end{figure}

{\bf Example:} Let $\llb X,Y\rrb = \{(0,0), (0,1), (1,1)\}$. As $\llb X,Y\rrb\neq \llb X\rrb\times\llb Y\rrb$ = $\{0,1\}^2$, 
$X$ and $Y$ are related. However, every pair of points in  $\llb X,Y\rrb$ is taxicab-connected, so $\mcT[X;Y]$ has only one set,
 $\llb X,Y\rrb$, and $\mrIs[X;Y]\stackrel{\mathrm{Cor.}\ \ref{taxicabcor}}{=}0$. 
See also Figure \ref{zeroinffig} for other examples.

This conservatism might suggest that $\mrIs$  
could  be derived from Shannon information via a {\em variational principle}, i.e. 
as  
\[
\inf\left\{ \mrI[X;Y]: \  X,Y \mbox{ rv's with given support }\llb X,Y\rrb\right\}.
\]
However, such an approach would be {\em too} conservative, since the infimum can be zero
even when the maximin information is strictly positive.   
A formal proof of this is not given due to space constraints, but a sketch of the argument follows.
Let $q$ be a (suitably well-behaved) joint probability density function (pdf) 
that is strictly positive on the Lebesgue measurable support $\llb X,Y\rrb$ 
of Figure 3(b) and  that has finite Shannon information.
Pick a point $x',y'$ in the interior of the support
and for any $\eps$ and sufficiently small $r>0$,   
let $X,Y$ be rv's with joint  pdf
$p_{X,Y}=(1-\eps)u_{x',y',r} + \eps q$,
where $u_{x',y',r}$ 
is a uniform pdf $u_{x',y'}$ on 
a square of dimension $r>0$ centred at $(x',y')$.
Observe that if $(r,\eps) = (0,0)$, the joint pdf becomes a unit delta function centred at $(x',y')$,
which automatically yields  zero mutual information.
As $\mrI[X;Y]$ must vary continuously with $\eps,r\geq 0$,
it follows that $\mrI[X;Y]\to 0$
as $(\eps,r)\to (0,0)$. 
The nonnegativity of Shannon information then implies that the infimum above must be zero,
but the maximin information  remains 1.

\section{Channels and Capacity}
\label{channelsec}
In this section, a connection is made between maximin information 
and the problem of transmission over an erroneous, discrete-time channel.

\subsection{Stationary Memoryless Uncertain Channels}

Let $\mbX^\infty$ be the space of all $\mbX$-valued, discrete-time functions
$x:\WW\to\mbX$. An {\em uncertain (discrete-time) signal} $X$ is a mapping
 from the sample space $\Ome$ to some function space  $\mcX\subseteq \mbX^\infty$ of interest.
Confining this mapping to any time $t\in\WW$ yields an uncertain variable (uv), denoted $X(t)$. 
The  signal segment $\left ( X(t)\right )_{t=a}^b$ is denoted $X(a:b)$.  
As with uv's, the dependence on $\ome\in\Ome$ will not usually be indicated:
thus the statements $X\in\mbA$ and  $X(t)=x(t)$ 
mean that $X(\ome)\in\mbA$ and  $X(\ome)(t)=x(t)$ respectively.
Also note that $\llb X\rrb$ here is a subset of the function space $\mcX$.  

A nonstochastic parallel of the standard notion of 
a {\em stationary memoryless channel} in communications can be defined as follows:
\begin{dfn}
\label{channeldfn}
Given an {\em input  function space}  $\mcX\subseteq\mbX^\infty$ and  a set-valued {\em transition function} $\mbT:\mbX\to 2^{\mbY}$,
a {\em  stationary memoryless uncertain channel}
maps any uncertain input signal $X$ with range $\llb X\rrb\subseteq\mcX$ 
to an  uncertain output signal $Y$
so that   
\bea 
\llb  Y(0:t)| x(0:t)\rrb   
&=&   \mbT(x(0))\times\cdots\times  \mbT(x(t)),\nn\\
& & 
\  \  x(0:t)\in\llb X(0:t)\rrb, t\in\WW.  
\label{defchannel}
\eea 
The set-valued {\em reverse transition function} $\mbR:\mbY\to 2^{\mbX}$ of the channel is 
\beq
\mbR(y):=\{  x\in\mbX: \mbT(x)\ni y\}, \ \  y\in\mbY. 
\label{defR}
\eeq 
\end{dfn}


{\bf Remarks:}
The set-valued map $\mbT$ here plays the role of  
a time-invariant transition probability matrix or kernel in communications theory.
The input function class $\mcX$ is included to handle possible constraints such as limited time-averaged transmission
power  or input run-lengths, though in the rest of this paper  $\mcX$ is taken as $\mbX^\infty$. 

The definition above implicitly assumes no feedback from the receiver
back to the transmitter. If such feedback is present then by arguments similar to Massey's \cite{massey},
a more general definition must be used  - see \cite{nairCDC12}.


The following lemma shows that the conditional range of the input sequence given an output sequence  
is defined by the  reverse transition function and the unconditional input range.
\begin{lem}
\label{invlem}
Given a  stationary memoryless uncertain channel (Definition \ref{channeldfn}) 
with   
 reverse transition function $\mbR$ (\ref{defR}),   
\beq   
\llb  X(0:t)|y(0:t)\rrb     
= \llb  X(0:t)\rrb   \cap \underbrace{\prod_{i=0}^t\mbR\left (y(i)\right )}_{=:\mbR\left (y(0:t)\right )}, \ \  
 \  y(0:t)\in\mbY^{t+1}
\label{invchannel}
\eeq  
and for any valid pair $X,Y$ of uncertain input and output signals.
\end{lem}

{\em Proof:}
See appendix  \ref{invlempf}. $\Box$

The largest information rate across a channel is formally defined as follows:
\begin{dfn}\label{Rsdfn}
The {\em peak maximin information rate} of  a stationary memoryless uncertain channel 
 (Definition \ref{channeldfn})
is
\beq
 R_* :=  
\sup_{t\in\WW, X: \llb  X\rrb   \subseteq\mcX}
\frac{\mrIs[  X(0:t);Y(0:t)]   }{t+1}, 
\label{defRs}
\eeq
where $\mcX$ is the input function space and $Y$ is the uncertain output signal
yielded by  the uncertain input signal $X$.
\end{dfn}

$\diamondsuit$

It can be shown that the term under the supremum over time on the RHS is 
 super-additive.  
A standard result called {\em Fekete's lemma}  then states that
 the supremum over time on the RHS of (\ref{defRs}) is achieved in the limit as $t\to\infty$. 
This leads immediately to the following
identity: 
\begin{lem}
\label{simplelem}
For any stationary  memoryless uncertain channel  (Definition \ref{channeldfn}),
the peak maximin information rate $ R_*$ (\ref{defRs}) satisfies 
\beq 
 R_* = 
\lim_{t\to\infty}\sup_{X: \llb  X\rrb   \subseteq\mcX}
\frac{\mrIs[  X(0:t);Y(0:t)]   }{t+1}, \label{defRs2}
\eeq
where  $\mcX$ is the input function space and  $Y$ is the uncertain output signal yielded by $X$.
\end{lem}

$\nabla$

\subsection{Zero-Error Capacity}

It is next shown how $ R_*$ relates to  the concept of {\em zero-error capacity} $C_0$
\cite{shannonTIT56,kornerTIT98}, which Shannon introduced after its more famous sibling the {\em (ordinary) capacity} $C$ \cite{shannon}. 
As described in section  \ref{introsec},  the zero-error capacity of a stochastic channel is defined as the largest
average block-coding bit-rate at which input ``messages'' can be transmitted while ensuring 
that the probability of a decoding error is exactly zero (not just arbitrarily small, as with the usual capacity).  
It is well known that $C_0$ does not depend on the probabilistic nature of the channel, in
the sense that the specific values of the nonzero transition probabilities play no role.
This suggests that  $C_0$ ought to be defineable using the nonstochastic framework of this paper.

To see this, observe that a  length-$(t+1)$ zero-error block code may be represented as a finite set  $\mbF\subseteq\mbX^{t+1}$,
where each codeword $f\in\mbA$ corresponds to a distinct ``message''.
The average coding rate is  thus $(\log|\mbF|)/(t+1)$ bits/sample,  
under the constraint that any received output sequence $y(0:t)$ corresponds to  at most  one possible $f$.
In other words  $\  t\in\WW$,
a set $\mbF\subseteq\mbX^{t+1}$ of codewords is valid iff for each possible channel output sequence $y(0:t)\in\mbY^{t+1}$, 
$|\mbF\cap\mbR(y(0:t))|\leq 1$.
Thus  the zero-error capacity may be defined operationally as 
\beq
C_0 := 
\sup_{t\in\WW,\mbF\in \bar{\mcF}(\mbX^{t+1})}
\frac{\log|\mbF|}{t+1}   
=\lim_{t\to\infty}\sup_{\mbF\in \bar{\mcF}(\mbX^{t+1})}
\frac{\log|\mbF|}{t+1},   
\label{defC0}
\eeq 
where the limit again follows from superadditivity and   
\bea
\lefteqn{\bar{\mcF}(\mbX^{t+1})
:=}\nn\\
& & \left\{\mbF\in\mcF(\mbX^{t+1}): \forall  y(0:t)\in\mbY^{t+1}, |\mbF\cap\mbR(y(0:t))|\leq 1  \right\},
\label{defFbar}
\eea
with $\mcF(\mbX^{t+1})$  the family of all finite subsets of $\mbX^{t+1}$ and 
$\mbR$, the reverse block transition function (\ref{invchannel}). 

The main result of this section shows that $C_0$ admits an intrinsic characterization in terms of
 maximin information theory:
\begin{thm}[$C_0$ via Maximin Information] 
\label{mainthm} 
For any stationary memoryless uncertain channel with input function space $\mcX=\mbX^\infty$ (Definition \ref{channeldfn}), 
the peak maximin information rate $R_*$ 
(Definition \ref{Rsdfn}) equals the zero-error capacity $C_0$ (\ref{defC0}).
\end{thm}  

{\em Proof:}
As $\llb X(0:t)|Y(0:t)\rrb_*$ is a partition of $\llb X(0:t)\rrb$, 
\bea
\lefteqn{|\llb X(0:t)|Y(0:t)\rrb_*|}\nn\\ 
&=&\sup_{\mbF\in \mcF\llb X(0:t)\rrb:\forall \mbC\in \llb X(0:t)|Y(0:t), |\mbF\cap\mbC|\leq 1 \rrb_*}|\mbF|
\nn\\
&\stackrel{(\ref{overlapsets})}{\leq} & \sup_{\mbF\in \mcF\llb X(0:t)\rrb: \forall \mbB\in \llb X(0:t)|Y(0:t), |\mbF\cap\mbB|\leq 1 \rrb}|\mbF|
\nn\\
&\stackrel{(\ref{invchannel})}{=} & \sup_{\mbF\in \mcF\llb X(0:t)\rrb: \forall y(0:t)\in \mbY^{t+1}, |\mbF\cap\mbR(y(0:t))|\leq 1}|\mbF|
\nn\\
&\stackrel{(\ref{defFbar})}{\leq} & \sup_{\mbF\in \bar{\mcF}(\mbX^{t+1})}|\mbF|
\stackrel{(\ref{defC0})}{\leq}
2^{C_0(t+1)}.
\label{Fbnd}\\
\Rightarrow  R_* &\stackrel{(\ref{Isimple}),(\ref{defRs})}{\leq}
C_0. 
\label{Rsbnd}
\eea

It is next shown that $\forall t\in\WW$, $\exists$ a uv $X(0:t)$ for which (\ref{Fbnd}) is an equality.
For any $\mbF\in \bar{\mcF}(\mbX^{t+1})$ (\ref{defFbar}),
let $X(0:t)$ be a surjection from $\Ome\to\mbF$.\footnote{As in the mutual-information characterization of Shannon capacity, it is implicit that the underlying
sample space $\Ome$ is  infinite, so that such a surjection always exists for each $t\in\WW$.} 
Then no point in $\llb X(0:t)\rrb=\mbF$ is overlap connected (Definition \ref{overlapdefn}) with any other,
since at least one of the conditional ranges $\llb X(0:t)|y(0:t)\rrb$ overlap-connecting them would then  
have 2 or more distinct points; this is impossible by (\ref{invchannel}) and (\ref{defFbar}).
Thus the overlap partition  $\llb X(0:t)|Y(0:t)\rrb_*$ (Lemma \ref{overlaplem}) 
of $\llb X(0:t)\rrb$ 
is a family of $|\llb X(0:t)\rrb|=|\mbF|$ singletons, comprising the individual points of $\llb X(0:t)\rrb=\mbF$.

If  $\bar{\mcF}(\mbX^{t+1})$ has a set $\mbF^*$ of maximum cardinality,
then choosing $\mbF=\mbF^*$ forces the LHS of (\ref{Fbnd}) to coincide with the RHS.
Otherwise, the RHS of (\ref{Fbnd}) will be infinite and $\mbF$ may be chosen to have arbitrarily large cardinality,
again yielding equality in (\ref{Fbnd}), by (\ref{Isimple}).
This achieves equality in (\ref{Rsbnd}). 
$\Box$ 

{\bf Remarks:} This result 
shows that the largest average bit-rate that can be transmitted across a stationary memoryless uncertain channel
with errorless decoding 
coincides with the largest average maximin information rate across it.
This parallels Shannon's channel coding theorem for stochastic memoryless channels and 
arguably makes      
 $\mrIs$  more relevant for problems involving communication than other nonstochastic information indices.

It must be noted that ensuring exactly zero decoding errors is a stringent
 requirement and is impossible over many common channels, such as the the binary symmetric, binary erasure
and additive white Gaussian noise channels, 
which have $C_0=0$. However,  a number of channels  are known to possess nonzero $C_0$,
 such as the pentagon and additive bounded noise channels.
Zero-error capacity is also an object of study in graph theory, where it is related to the
{\em clique number}.
 See \cite{kornerTIT98} for a comprehensive survey of the literature on $C_0$.



 
\section{State Estimation of Linear Systems over Erroneous Channels} 
\label{stateestsec}

In this section,  maximin information 
is used to study the problem of estimating the states of 
a linear time-invariant (LTI) plant via a stationary memoryless uncertain channel (Definition \ref{channeldfn}),
without channel feedback. First, some related prior work is discussed. 

\subsection{Prior Work}

In the case where the channel is an errorless digital bit-pipe, the state estimation problem
is formally equivalent to feedback stabilization with control inputs known to both encoder and  decoder.
The central result in this scenario is the so-called ``data rate theorem'', which  states that  
the estimation error or plant state
 can be stabilized or taken to zero iff the sum $H$ of the log-magnitudes of the
unstable eigenvalues of the system is less than the channel bit-rate. 
This condition holds in both deterministic and probabilistic settings,
and  under different notions of convergence or stability, 
e.g. uniform, $r$th moment or almost surely (a.s.) \cite{wong2,wong3,hespanha,baillieul2,tatikondaTACJuly04,nair13,nair16}. 
See also \cite{you11} for recent work on quantized estimation of stochastic LTI systems.   

However, if transmission errors 
 occur, then the stabilizability and estimation conditions become  
highly dependent on the setting and objective, leading to a variety of different criteria.
For instance, given a stochastic discrete memoryless channel (DMC) 
and a noiseless LTI system with random initial state,
a.s. convergence of the state or estimation error to zero is possible 
if and (almost) only if the ordinary channel capacity $C\geq H$;
this was proved for digital packet-drop channels with acknowledgements  
in \cite{tatikondaTACJuly04b}, and for general DMC's with or 
without channel feedback in \cite{matveevSICON07}.
The same result also holds for asymptotic stabilizability  
via an additive white Gaussian 
noise channel \cite{braslavskyTAC07}, with no channel feedback.
See also \cite{comoSICON10} for bounds on mean-square-error convergence rates
for state estimation over stochastic DMC's, without channel feedback.
 

Suppose next that additive stochastic noise perturbs the plant 
 and the objective is to  bound the $r$th moment of the states or estimation errors.
Assuming channel feedback, bounded noise and scalar states,
the achievability of this goal is determined by   
the {\em anytime capacity} of the channel \cite{sahaiTIT06}. 
Other related articles are \cite{martinsTAC06,mineroTAC09,yukselTAC11}
 - the first two consider moment stabilization over errorless channels with randomly varying bit-rates 
known to both transmitter and receiver, and the last studies mean-square stabilization 
via DMC's with no channel feedback. 
See also the recent papers \cite{ostrovskyTIT09,sukhavasi12} for explicit constructions of
error-correcting codes for control. 

For the purposes of this section, the most relevant prior work
is  \cite{matveevIJC07} (see also \cite{matveevBook}),
in which the channel is modelled as a stochastic DMC,
and the plant is LTI with  random initial state 
but is perturbed by additive nonstochastic bounded disturbances.
It was shown that if channel feedback is absent, then 
a.s. uniformly bounded estimation errors are possible iff  $H< C_0$, 
the zero-error capacity \cite{shannonTIT56} of the channel.
However, under perfect channel feedback    the necessary and sufficient
condition becomes $H< C_{0\mrf}$, the {\em zero-error feedback capacity}  defined in \cite{shannonTIT56};
the same criterion applies if  the goal is to stabilize the plant
states in the a.s. uniformly bounded sense, with or without channel feedback.
As $C_0$ and $C_{0\mrf}$ are (often strictly) less than $C$,  
both these conditions are more restrictive than for plants 
with stochastic or no process noise,
 even if the disturbance bound is arbitrarily small.
In rough terms, the reason for the increased strictness is 
that nonstochastic disturbances do not enjoy a law of large numbers that
averages them out in the long run.  
As a result it becomes crucial for no decoding errors to occur in the channel, not just for their average probability    
to be arbitrarily small. 
This important result was proved using probability theory,  a law of large numbers and volume-partitioning arguments, 
but no information theory.  

The scenarios considered in this section are similar to \cite{matveevIJC07},  
with the chief difference being that 
 that  neither the initial state nor the erroneous channel  
are modelled stochastically here. As a consequence,  probability and the law of large numbers cannot be employed in the analysis.
Instead,  maximin information is applied to 
yield necessary conditions that are then be shown to be tight (Thms. \ref{capthm} and \ref{capthm2}).
Only state estimation without channel feedback is considered here,
since the maximin-information theoretic analysis of
systems with feedback is significantly different - see 
\cite{nairCDC12} for some preliminary results.
 
In what follows,  $\|.\|$ denotes either the maximum norm  
on a finite-dimensional real vector space or the matrix norm it induces, 
and $\mbB_l(x)$ denotes the corresponding  $l$-ball $\{y:\|y-x\|\leq l\}$ centered at $x$.

\subsection{Disturbance-Free LTI Systems}
\label{distfreesec}

Consider an undisturbed linear time-invariant (LTI) system
\bea 
X(t+1) &=& AX(t) \ \ \in\RR^n, 
\label{dyn}
\\
Y(t) &=& GX(t) \ \ \in\RR^p,  
\ \ \  t\in\WW, \label{meas}
\eea
where the initial state $X(0)$ is an uncertain variable (uv). 
The output signal $Y$ is causally encoded via an operator 
$\gamma$ as 
\beq
S(t)=\gamma\left (t,Y(0:t)\right )\in\mbS, \ \ \  t\in\WW.
\label{St}
\eeq  
Each symbol $S(t)$ is then transmitted over a stationary memoryless uncertain channel 
with set-valued transition function $\mbS\mapsto 2^{\mbQ}$ and input function space $\mbS^\infty$ (Definition \ref{channeldfn}),
 yielding a received symbol $Q(t)\in\mbQ$.
Note that the encoder is told nothing about the values of these received symbols,  i.e. there is no channel feedback. 
These symbols are used to produce a causal prediction $\Xh(t+1)$ of $X(t+1)$
by means of another operator $\eta$ as  
\beq
\Xh(t+1)\equiv \eta(t,Q(0:t))\in\RR^n, \ \ \  t\in\WW, \ \ \Xh_0=0.
\label{Xh}
\eeq
Let $E(t):=X(t)-\Xh(t)$ denote the prediction error.

The pair $(\gamma,\eta)$ is called a {\em coder-estimator}. 
Such a pair is said to yield $\vro${\em-exponential  
uniformly bounded errors} if  
for any uv $X(0)$ with range $\subseteq \mbB_l(0)$,  
\beq
\sup_{t\in\WW, \ome\in\Ome}\vro^{-t}\|E(t)\|
\equiv\sup_{t\in\WW} \sup\left\llb  \vro^{-t}\|E(t)\| \right\rrb <\infty, 
\label{unifbnd}
\eeq 
where $l,\vro>0$ are specified parameters.
 If  the stronger property 
\beq
\lim_{t\to\infty}\sup_{\ome\in\Ome}\vro^{-t}\|E(t)\|
\equiv\lim_{t\to\infty} \sup\left\llb  \vro^{-t}\|E(t)\| \right\rrb =0 
\label{unif}
\eeq 
holds, then $\vro${\em-exponential  
uniform convergence} is said to be achieved.

Impose the following assumptions:
\begin{description}
\item[DF1:] The pair $(G,A)$ in (\ref{dyn})--(\ref{meas}) is observable.
\item[DF2:]
For every $t\in\WW$, 
the channel output sequence $Q(0:t)$ (Definition \ref{channeldfn}) is conditionally unrelated (Definition \ref{unrelateddfn}) 
with initial state $X(0)$, given the channel input sequence $S(0:t)$;
i.e. $ X(0)\leftrightarrow S(0:t)\leftrightarrow Q(0:t)$.
\item[DF3:] The convergence parameter $\vro$ of (\ref{unifbnd})--(\ref{unif}) is strictly smaller than the spectral radius of $A$.
\end{description}

{\bf Remarks:} 
Condition DF1 can be relaxed to requiring the observability  of $A$ 
on the invariant subspace  corresponding to eigenvalues greater than or equal to $\vro$ in magnitude.  
Assumption DF2 basically states that the channel outputs can depend on the initial state only
via the channel inputs. 
Condition DF3 entails negligible loss of generality,
since if $\vro$ were to exceed the largest plant eigenvalue magnitude $|\lambda_{\max}|$, 
then  the trivial estimator $\Xh(t)=0$ would achieve (\ref{unif}) and communication would not be needed.\footnote{The 
case $\vro =|\lambda_{\max}|$ introduces technicalities that can be handled by modifying to the arguments below;
for the sake of conciseness it is not explicitly treated here.}

The main result of this subsection is given below:
\begin{thm}
\label{capthm}
Consider the linear time-invariant system (\ref{dyn})--(\ref{meas}), with plant matrix $A\in\RR^{n\times n}$, uncertain initial state $X(0)$
and outputs that are coded and estimated (\ref{St})--(\ref{Xh}) without channel feedback,
via a stationary memoryless uncertain channel 
 (Definition \ref{channeldfn}) with
zero-error capacity $C_0\geq 0$ (\ref{defC0}). 
Let $\lambda_1,\ldots ,\lambda_n$ be the eigenvalues of $A$
and suppose that Assumptions DF1--DF3 hold.

If there exists a coder-estimator that yields  
 $\vro$-exponential uniformly bounded estimation errors   
(\ref{unifbnd}) 
with respect to a nonempty $l$-ball $\mbB_l(0)\subset\RR^n$    
of initial states, then
\beq
C_0 \geq   
\sum_{i\in[1:n]:|\lambda_i|\geq \vro}\log\left |\frac{\lambda_i}{\vro}\right | =:H_\vro.
\label{defHro}
\eeq
 
Conversely, if the inequality in (\ref{defHro}) holds strictly, 
then a coder-estimator without channel feedback can be constructed to yield 
$\vro$-exponential uniform convergence  
(\ref{unif}) 
on any initial-state $l$-ball.
\end{thm}


\subsubsection{Proof of Necessity}
\label{necsec}
The necessity of (\ref{defHro}) is established first. 
Without loss of  generality,
let the state coordinates be chosen so that $A$ is in  
{\em real Jordan canonical form} (see e.g. \cite{horn}, Theorem  3.4.5), 
i.e. it consists of $m$ square blocks on its diagonal, with the $j$th block $A_j\in\RR^{n_j\times n_j}$ having either
identical real eigenvalues  or 
identical complex eigenvalues and conjugates for each $j\in[1:m]$.
Let the blocks be  ordered by descending eigenvalue magnitude.
For any $j\in [1:m]$, let $X_j(t)\in\RR^{n_j}$ comprise those components of $X(t)$ governed by the $j$th real Jordan block $A_j$, 
and  let $E_j(t),\Xh_j(t)\in\RR^{n_j}$ consist of the corresponding components of $E(t)$ and $\Xh(t)$, respectively.

Let $d\in[0:n]$ denote the number of eigenvalues with magnitude $> \vro$, including repeats. 
Pick arbitrary
$\tau\in\NN$ and  
\beq
\eps\in \left ( 0, 1-\max_{i:|\lambda_i|>\vro}\frac{\vro}{|\lambda_i|}\right ),
\label{eps}
\eeq
and then divide  
the interval $[-l,l]$  on the $i$th axis into 
\beq
k_i:= \left\lfloor \left |\frac{(1-\eps)\lambda_i}{\vro}\right |^\tau \right\rfloor
\label{defki}
\eeq
equal subintervals of length $2l/k_i$,
for each $i\in[1:d]$.
Denote the midpoints of the subintervals so formed by $p_i(s)$, $s=1,\ldots , k_i$, 
and inside each subinterval construct an interval $\mbI_i(s)$ centred at $p_i(s)$ but of shorter length  $l/k_i$.
Define a  hypercuboid family  
\beq
\mcH:=\left\{\left (\prod_{i=1}^{d}\mbI_i(s_i)\right )\times [-l,l]^{n-d}: s_i\in [1:k_i], i\in[1:d]\right\}
\label{defmcH}
\eeq 
and observe that any two hypercuboids $\in\mcH$ are 
 separated by a distance  of $l/k_i$ along the $i$th axis for each $i\in[1:d]$.
Set the initial state range $\llb X(0)\rrb = \bigcup_{\mbH\in\mcH}\subset\mbB_l(0)$.


As  $\llb  E_j(t)\rrb   \supseteq \llb  E_j(t)|q(0:t-1)\rrb$, 
\bea 
\lefteqn{\dia\llb E_j(t)\rrb
 \geq   \dia\left\llb E_j(t) |q(0:t-1)\right\rrb} 
\nn\\
&=&  \dia\left\llb A^t_jX_{j}(0)- \eta_{j}\left (t,q(0:t-1)\right ) |q(0:t-1)\right\rrb 
\nn\\
&=&  \dia\left\llb A^t_jX_{j}(0) |q(0:t-1)\right\rrb 
\label{trans}\\
&\equiv & \sup_{u,v\in\left\llb X_{j}(0) |q(0:t-1)\right\rrb }
\|A^t_j(u-v)\|
\nn\\
&\geq &   \sup_{u,v\in\left\llb X_{j}(0) |q(0:t-1)\right\rrb }
\frac{\|A^t_j(u-v)\|_2}{\sqrt{n}} 
\nn\\
&\geq & \sup_{u,v\in\left\llb X_{j}(0) |q(0:t-1)\right\rrb }
\frac{\sigma_{\min}(A^t_j)\|u-v\|_2}{\sqrt{n}} 
\label{minsing}
\\
&\geq & \sup_{u,v\in\left\llb X_{j}(0) |q(0:t-1)\right\rrb }
\frac{\sigma_{\min}(A^t_j)\|u-v\|}{\sqrt{n}} 
\nn\\
&\equiv &  \sigma_{\min}(A^t_j)\frac{\dia\left\llb X_{j}(0) |q(0:t-1)\right\rrb}{\sqrt{n}},
\nn\\
& & \  t\in\WW, \  q(0:t-1)\in \llb Q(0:t-1)\rrb,
\label{minsing2}
\eea
where diam$(\cdot)$ denotes set diameter under the maximum norm; (\ref{trans}) holds since translating a set in a normed space does not change its diameter;
$\|\cdot\|_2$ denotes Euclidean norm;
and 
 $\sigma_{\min}(\cdot) $ denotes smallest singular value. 

Now, an asymptotic identity of Yamamoto 
states that 
$\lim_{t\to\infty}\left (\sigma_{\min}  (A^t_j)\right )^{1/t} =   |\lambda_{\min}(A_j)|$,
 where $\lambda_{\min}(\cdot)$ denotes smallest-magnitude eigenvalue (see e.g. \cite{horn2}, Thm 3.3.21).
As there are only finitely many blocks $A_j$,  $\exists t_\eps\in\WW$ s.t. 
\beq
\sigma_{\min}(A^t_j)\geq \left (1-\frac{\eps}{2}\right )^t |\lambda_{\min}(A_j)|^t,
\ \ \  j\in[1:m], \ t\geq  t_\eps.
\label{yambnd}
\eeq
In addition, for any region $\mbK$ in a normed vector space,
\bea
\dia(\mbK)&\equiv&\sup_{u,v\in\mbK}\|u-v\|
\leq \sup_{u,v\in\mbK}\|u\| +\|v\|
\nn\\
&=& 2\sup_{u\in\mbK}\|u\|. 
\label{diambnd}
\eea
By (\ref{unifbnd}),  there then exists $\phi>0$ such that   
\bea
\lefteqn{\phi\vro^t  \geq  \sup\llb \|E(t)\| \rrb}\nn\\
& \geq &\sup\llb \|E_j(t)\| \rrb
\stackrel{(\ref{diambnd})}{\geq} 
 0.5 \dia\llb E_j(t)\rrb 
\nn\\
& \stackrel{(\ref{yambnd}),(\ref{minsing2})}{\geq}&  
 \left | \lp 1-\frac{\eps}{2}\rp\lambda_{\min}(A_j)\right |^t \frac{\dia\left\llb X_{j}(0) |q(0:t-1)\right\rrb}{2\sqrt{n}},
\nn\\
& & 
\  j\in[1:m], \ t\geq  t_\eps. 
\label{yamamoto}
\eea

For some $\tau\in\NN$,  
the hypercuboid family $\mcH$  (\ref{defmcH})  is an $\llb X(0)| Q(0:\tau-1)\rrb$-overlap isolated partition (Definition \ref{overlapdefn}) of $\llb X(0)\rrb$.
To see this, suppose in contradiction that $\exists \mbH\in\mcH$  that is overlap connected  in  $\llb X(0)| Q(0:\tau-1)\rrb$  
with another hypercuboid in $\mcH$.
Then there would exist a set $\llb X(0): q(0:\tau-1)\rrb$ containing a point $u\in\mbH$ and a point $v$ in some   
$\mbH'\in\mcH\setminus\{\mbH\}$.      
Thus $u_j,v_j\in \llb X_j(0)| q(0:\tau-1)\rrb$,
implying  
\bea
\|u_j-v_j\| &\leq &\dia \llb X_j(0)| q(0:\tau-1)\rrb \nn\\
&\stackrel{(\ref{yamamoto})}{\leq}& 
 \frac{2\sqrt{n}\phi \vro^\tau}{\left | (1-\eps/2)\lambda_{\min}(A_j)\right |^\tau},
\nn\\
& & \ j\in[1:m], \ \tau\geq  t_\eps.
\label{distbnd1}
\eea
However, by construction  any two hypercuboids $\in\mcH$ are disjoint
and separated by a distance  of at least $l/k_i$ along the $i$th axis for each $i\in[1:d]$.
Thus if $A_j$ is the real Jordan block corresponding to some eigenvalue $\lambda_i$,  $i\in [1:d]$,
then 
\bean 
\|u_j-v_j\| &\geq & \frac{l}{k_i}  \stackrel{(\ref{defki})}{=}  
\frac{l}{\left\lfloor \left ((1-\eps)|\lambda_i|/\vro\right )^\tau \right\rfloor}
\\
&\geq &  \frac{l}{\left ((1-\eps)|\lambda_i|/\vro\right )^\tau }
 =  \frac{l\vro^\tau}{\left |(1-\eps)\lambda_{\min}(A_j)\right |^\tau },
\eean 
since all the eigenvalues of $A_j$ have equal magnitudes.
The RHS of this would exceed the RHS of (\ref{distbnd1}) when
$\tau\geq  \max(t_\eps,t')$ is sufficiently large that $\left (\frac{1-\eps/2}{1-\eps}\right )^\tau > 2\sqrt{n}\phi/l$,
yielding a contradiction.

As  $\mcH$ is an $\llb X(0)| Q(0:\tau-1)\rrb$-overlap isolated partition of $\llb X(0)\rrb$  for sufficiently large $\tau$, 
\bea
2^{\mrIs[X(0); Q(0:\tau-1)]}& \stackrel{(\ref{Isimple})}{=} & \left |\llb X(0)|Q(0:\tau-1)\rrb_*\right | \stackrel{(\ref{maxcard})}{\geq}  |\mcH| 
\nn\\
& = & \prod_{i=1}^d k_i \stackrel{(\ref{defki})}{=} \prod_{i=1}^d \left\lfloor \left |\frac{(1-\eps)\lambda_i}{\vro}\right |^\tau \right\rfloor
\nn\\
&\geq & \prod_{i=1}^d 0.5 \left |\frac{(1-\eps)\lambda_i}{\vro}\right |^\tau  
\label{halfbnd}\\
&= &  \frac{(1-\eps)^{d\tau} \left |\prod_{i=1}^d\lambda_i\right |^\tau }{2^d \vro^{d\tau}},
\label{Islbnd}
\eea
where (\ref{halfbnd}) follows from (\ref{eps}) and the inequality $\lfloor x\rfloor > x/2$, for every $x\geq 1$.
However,  since $X(0)\leftrightarrow S(0:\tau-1)\leftrightarrow Q(0:\tau-1)$ is a Markov uncertainty-chain (Definition \ref{markovdfn}), 
\bean  
\mrIs [  X(0); Q(0:\tau-1) ]   
& \stackrel{\mrm{Lem.}\ \ref{DPlem}}{\leq} & 
\mrIs [ S(0:\tau-1); Q(0:\tau-1)] \\
&\stackrel{\mrm{Def.}\ \ref{Rsdfn}}{\leq}& \tau R_*
\\
& \stackrel{\mrm{Thm.}\ref{mainthm}}{=}& \tau C_0. 
\eean 
Substituting this into the LHS of (\ref{Islbnd}), taking logarithms, dividing by $\tau$ and then letting $\tau\to\infty$ yields 
\[ 
 C_0  \geq  d\log(1-\eps) +\sum_{i=1}^d\log\left |\frac{\lambda_i}{\vro}\right |.
\]
As $\eps$ may be arbitrarily small, this establishes the necessity of (\ref{defHro}).

\subsubsection{Proof of Sufficiency}
\label{suffproof}
The sufficiency of (\ref{defHro}) is straightforward to establish.
Define new state and measurement vectors  $X'(t)=\vro^{-t} X(t)$ and $Y'(t)=\vro^{-t} Y(t)$, for every $t\in\WW$. 
In these new coordinates, the system equations (\ref{dyn})--(\ref{meas}) become
\bea 
X'(t+1) &=& (A/\vro)X'(t) \ \ \in\RR^n, 
\label{dynro}
\\
Y'(t) &=& GX'(t) \ \ \in\RR^p,  
\ \ \  t\in\WW. \label{measro}
\eea
By (\ref{defHro}) and   (\ref{defC0}),
 $\forall\delta\in  (0, C_0-H_\vro )$ 
$\exists t_\delta>0$ s.t. $\forall\tau\geq t_\delta$, $\exists$ a finite set $\mbF\subseteq\mbS^{\tau}$ with 
$\max_{q_0^{\tau-1}\in\mbQ^{\tau}}  |\mbF\cap\mbR(q_0^{\tau-1}) | = 1$ and 
\beq 
H_\vro <  C_0-\delta  \leq   (\log|\mbF|)/\tau. 
\label{bndHr}
\eeq
Down-sample  (\ref{dynro})--(\ref{measro})
by $\tau$ to obtain the  LTI system
\bea 
X'\left ((k+1)\tau\right ) &=& (A/\vro)^\tau X'(k\tau) \ \ \in\RR^n, 
\label{dynrotau}
\\
Y'(k\tau) &=& GX'(k\tau) \ \ \in\RR^p,  
\ \ \  k\in\WW. \label{measrotau}
\eea
Now,  $|\mbF|$ distinct codewords can be transmitted over the channel and decoded without error
once every $\tau$ samples. 
Furthermore  $\log|\mbF|\stackrel{(\ref{bndHr})}{>}\tau H_\vro =$  sum of the unstable eigenvalue log-magnitudes of $(A/\vro)^\tau$.  
By the ``data rate theorem'' (see e.g. \cite{wong2}), there then exists a coder-estimator for the LTI down-sampled
system  (\ref{dynrotau})--(\ref{measrotau}) that estimates the states of (\ref{dynrotau}) with errors 
$\|X'(k\tau)-\Xh'_k\|$ tending uniformly to 0.
For every $t\in\WW$, write $t=k\tau +r$ for some $k\in\WW$ and $r\in[0:\tau-1]$,
 and define an estimator  
\[
\Xh(t):=\vro^{k\tau+r}  A^r\Xh'_k. 
\]
Then 
\bean
\lefteqn{\vro^{-t}\sup_{\ome\in\Ome}\|X(t)-\Xh(t)\|}\\
&=& \vro^{-(k\tau+r)}\sup_{\ome\in\Ome}
\left \|\vro^{k\tau} A^r X'(k\tau) -  \vro^{k\tau}  A^r\Xh'_k\right\|  
\\
&\leq & \vro^{-r}\|A^r\|\sup_{\ome\in\Ome} \left\|X'(k\tau) -   \Xh'_k\right\| 
\\
&\leq & \max_{r\in[1:\tau-1]} 
\left\{\vro^{-r}\|A^r\|\right\} \sup_{\ome\in\Ome}\left\|X'(k\tau) -   \Xh'_k\right\| 
\to 0
\eean
as $t$, and hence $k\equiv \lfloor t/\tau\rfloor$, tend to $\infty$.

\subsection{LTI Systems with Disturbances}
\label{distsec}

The results and techniques of the previous subsection
can be readily adapted to analyze systems with disturbances.  
Suppose that, instead of (\ref{dyn})--(\ref{meas}),
the plant state and output equations are
\bea 
X(t+1) &=& AX(t) +V(t)\ \ \in\RR^n, 
\label{dyn2}
\\
Y(t) &=& GX(t) +W(t)\ \ \in\RR^p,  
\ \  t\in\WW, \label{meas2}
\eea
where the uncertain signals $V$ and $W$  represent additive process and measurement noise.
The objective is {\em uniform boundedness}, i.e. (\ref{unifbnd}) with $\vro=1$. 
Make the following assumptions:
\begin{description}
\item[D1:] The plant dynamics (\ref{dyn2}) are strictly unstable, i.e. the matrix $A$ has spectral radius strictly larger than 1.
\item[D2:] The uncertain noise signals $V$ and $W$ are uniformly bounded, i.e. $\exists c>0$ s.t. 
all possible signal realizations $v\in\llb V\rrb$ and $w\in\llb W\rrb$  
have $\ell^\infty$-norms $\|v\|,\|w\|\leq c$.
\item[D3:] The zero sequence is a possible process and measurement noise realization, 
i.e. $0\in\llb V\rrb\cap\llb W\rrb$.  
\item[D4:] The initial state $X(0)$, $V$ and $W$ are mutually unrelated (Definition \ref{unrelateddfn}). 
\item[D5:] 
For every $t\in\WW$, 
the channel output sequence $Q(0:t)$ (Definition \ref{channeldfn}) is conditionally unrelated (Definition \ref{unrelateddfn}) 
with $\lp X(0),V(0:t-1),W(0:t)\rp$, given the channel input sequence $S(0:t)$,
i.e. $\lp X(0),V(0:t-1),W(0:t)\rp\leftrightarrow S(0:t)\leftrightarrow Q(0:t)$.
\end{description}

The following result holds:
\begin{thm}
\label{capthm2}
Consider a linear time-invariant plant (\ref{dyn2})--(\ref{meas2}), with plant matrix $A\in\RR^{n\times n}$, uncertain initial state $X(0)$,
and bounded uncertain signals $V$ and $W$ additively corrupting the dynamics and outputs respectively. 
Suppose the plant outputs are coded and estimated (\ref{St})--(\ref{Xh}) without feedback via a stationary memoryless uncertain channel 
 (Definition \ref{channeldfn}) having  
zero-error capacity $C_0\geq 0$ (\ref{defC0}), and assume conditions DF1 and D1--D5.

If there exists a coder-estimator (\ref{St})--(\ref{Xh})  yielding  
uniformly bounded estimation errors 
with respect to a nonempty $l$-ball $\mbB_l(0)\subset\RR^n$    
of initial states, 
then
\beq
C_0 \geq   
\sum_{i\in[1:n]:|\lambda_i|\geq 1}\log |\lambda_i |=: H,
\label{defH}
\eeq
where $\lambda_1,\ldots ,\lambda_n$ are the eigenvalues of $A$.
 
Conversely, if (\ref{defH}) holds as a strict inequality,
then a coder-estimator can be constructed to yield 
 uniform boundedness   
for any given $l$-ball of initial states.
\end{thm}

{\em Proof:}
Necessity is straightforward. If a coder-estimator
achieves uniform boundedness, then this uniform bound is not exceeded
if the uncertain disturbances  
are realized as the zero signal, which by hypothesis is an element of both  $\llb V\rrb$ and $\llb W\rrb$.
By unrelatedness $\llb X(0)|V=0,W=0\rrb=\llb X(0)\rrb$, so  the initial state range is unchanged.
Furthermore, condition D5 implies $ X(0)\leftrightarrow S(0:t)\leftrightarrow Q(0:t)$, i.e. condition DF2. 
As uniform boundedness is just $\vro$-exponential uniform boundedness
with $\vro=1$ (\ref{unifbnd}), Theorem  \ref{capthm} applies immediately to yield
(\ref{defH}).

The sufficiency of (\ref{defH}) is established next. 
By (\ref{defH}) and   (\ref{defC0}),
 $\forall\delta\in  (0, C_0-H )$ 
$\exists t_\delta>0$ s.t. $\forall\tau\geq t_\delta$, $\exists$ a finite set $\mbF\subseteq\mbS^{\tau}$ with 
$\max_{q_0^{\tau-1}\in\mbQ^{\tau}}  |\mbF\cap\mbR(q_0^{\tau-1}) | = 1$ and 
\beq 
H <  C_0-\delta  \leq   (\log|\mbF|)/\tau. 
\label{bndH}
\eeq
Down-sample  (\ref{dyn2})--(\ref{meas2})
by $\tau$ to obtain the  LTI system
\bea 
X\left ((k+1)\tau\right ) &=& A^\tau X'(k\tau) +V'_\tau(k)\ \ \in\RR^n, 
\label{dyntau}
\\
Y(k\tau) &=& GX(k\tau) +W(k\tau)\  \in\RR^p,  
\ \  k\in\WW, \label{meastau}
\eea
where the accumulated noise term $V'_r(k):=\sum_{i=0}^{r}A^{\tau-1-i} V(k\tau+i)$ 
can be shown to be uniformly bounded over $k\in\WW$ for each $r\in[0:\tau-1]$.
Now,  $|\mbF|$ distinct codewords can be transmitted over the channel and decoded without error
once every $\tau$ samples. 
Furthermore  $\log|\mbF|\stackrel{(\ref{bndH})}{>}\tau H =$  sum of the unstable eigenvalue log-magnitudes of $A^\tau$.  
By the ``data rate theorem'' for LTI systems with bounded disturbances controlled or estimated over errorless channels,
(see e.g. \cite{wong2,tatikondaTACJuly04,hespanha}), there then exists a coder-estimator for the LTI down-sampled
system  (\ref{dyn2})--(\ref{meas2}) that estimates its states with errors 
$X(k\tau)-\Xh_k$ uniformly bounded over $k\in\WW$. 

For every $t\in\WW$, write $t=k\tau +r$ for some $k\in\WW$ and $r\in[0:\tau-1]$,
 and define an estimator  
\[
\Xh(t):= A^r\Xh_k. 
\]
Then 
\bean
\lefteqn{\sup_{\ome\in\Ome}\|X(t)-\Xh(t)\| }\\
&=& \sup_{\ome\in\Ome}
\left \| A^r X(k\tau) +V'_r(k) -   A^r\Xh_k\right\|  
\\
&\leq & \|A^r\|\sup_{\ome\in\Ome} \left\|X(k\tau) -   \Xh_k\right\| 
+ \|V'_r(k)\|
\\
&\leq & \max_{r\in[1:\tau-1]} 
\left\{\|A^r\|\right\} \sup_{\ome\in\Ome}\left\|X(k\tau) -   \Xh_k\right\|
+ \max_{r\in[1:\tau-1]}\|V'_r(k)\|.
\eean
As the RHS is uniformly bounded over $k\in\WW$, the proof is complete.
$\Box$

\subsection{Discussion}

Like the results of Matveev and Savkin \cite{matveevIJC07} on LTI state estimation 
via an erroneous channel without feedback,  
Thms. \ref{capthm} and \ref{capthm2} involve the zero-error capacity of the channel.
In their formulation, the process and measurement noise are 
treated as  bounded unknown deterministic signals,  
but the channel and initial state are  modelled probabilistically.
The estimation objective is to achieve  estimation errors 
that, with probability (w.p.) 1, are uniformly  
bounded over  all admissible disturbances, 
and the necessity part of their result was proved with the aid of a law of large numbers.

The main aims of this section have been  to demonstrate firstly,
that statistical assumptions are not necessary to capture the
essence of this problem (modulo zero-probability events); and secondly,
that even with no probabilistic structure to exploit, 
information-theoretic techniques can be successfully applied, based on $\mrIs$. 
Although the channel and initial state here are modelled nonstochastically
and, furthermore, the estimation errors are to be bounded uniformly over all samples $\ome\in\Ome$,
not just w.p.1,  
the achievability criterion (\ref{defH}) of subsection  \ref{distsec} 
essentially recovers the earlier result.\footnote{The only difference is that the necessary condition here 
is not a strict inequality as in the earlier result, 
 because the proof technique here  relies on nulling the disturbances. 
A lengthier analysis that explicitly considers process noise effects would elicit 
a strict inequality; due to space constraints this is omitted.} 

In addition, unlike \cite{matveevIJC07} and Theorem  \ref{capthm2}, Theorem  \ref{capthm} 
 assumes no disturbances and
 concerns performance as measured by a specific convergence rate, 
not just bounded errors. 
The criterion (\ref{defHro}) agrees  with \cite{matveevIJC07} when $\vro=1$, 
but is more (less) stringent when $\vro <(>) 1$.
It applies when, for instance, the states of a possibly stable noiseless LTI plant are to be remotely estimated   
with errors decaying at or faster than a specified speed $\vro^t$.

 


\section{Conclusion}
In this paper  a formal framework for modelling nonstochastic  variables was proposed, 
leading to analogues of probabilistic ideas such as independence and  Markov chains. 
Using this framework,  the  concept of  maximin information was  introduced, and 
it was proved that the zero-error capacity  $C_0$ of a stationary memoryless uncertain channel 
coincides with the highest rate of maximin information across it.  Finally, 
maximin information was applied to the problem of reconstructing the states of
 a  linear time-invariant (LTI) system via such a channel.
Tight criteria involving $C_0$ were found for the achievability of  
  uniformly bounded and uniformly exponentially converging
estimation errors, without any statistical assumptions.  

An open question is whether maximin information can be used in the presence of feedback.
Two challenges present themselves.
Firstly,  the equivalence  between the problems of
state estimation and control in the errorless case is lost if channel errors occur, because
 the encoder does not necessarily know what the decoder received.
Secondly, from \cite{matveevIJC07,matveevBook} it is known that for both the problems 
of LTI state estimation with channel feedback
and LTI control, 
the relevant channel figure-of-merit  for achieving a.s. bounded estimation errors or states respectively
is its  {\em zero-error feedback capacity} $C_{0\mrf}$, 
which can be strictly larger than $C_0$ \cite{shannonTIT56}.

These issues suggest that nontrivial modifications of the techniques presented here may be required
to study feedback systems. 
Preliminary results concerning this problem are presented
in the conference paper \cite{nairCDC12}. 

\section*{Acknowledgements} 

The author acknowledges the helpful suggestions of the anonymous reviewers.


\bibliographystyle{IEEEtran}

\bibliography{IEEEabrv,rep1ref}

\appendices 

\section{Proof of Lemma  \ref{splitlem}}
\label{splitlempf}

By (\ref{subsplit}), it need only be established that the RHS of (\ref{split})
is contained in its LHS. 
Pick any realization $(y_1,\ldots , y_m)\in\llb  Y_1,\ldots , Y_m\rrb   $
and consider any element $x$ in the RHS of (\ref{split}).

Pick  any $(y_1,\ldots , y_m)\in\llb  Y_1,\ldots , Y_m\rrb   $
and any point $x\in$ the RHS. For every $i\in[1,\ldots ,n]$, 
$\exists \ome_i\in\Ome$ s.t. $X(\ome_i)=x$ and $Y_i(\ome_i)=y_i$,
so that $y_i\in\llb  Y_i|x\rrb   $.
By the conditional unrelatedness of $Y_1,\ldots , Y_i$ given $X$,
it follows that $(y_1,\ldots , y_m)\in \llb  Y_1,\ldots , Y_m|X\rrb   $.
That is,  $\exists \ome\in\Ome$ with $X(\ome)=x$ and $Y_i(\ome)=y_i$,
for each $i\in [1,\ldots, m]$. Thus $x\in\llb  X|y_1,\ldots , y_m\rrb   $,
implying that the the  RHS of (\ref{split}) is contained in the LHS.
By (\ref{subsplit}), the  LHS is also contained in the RHS,
establishing equality. 

\section{Proof of Lemma  \ref{overlaplem} (Unique Overlap Partition)}
\label{overlaplempf}

The first step is to establish the existence of an overlap partition.
For any $x\in \llb X\rrb$,  let $\mbO(x)$ be the set of all points
in $\llb X\rrb$ with which $x$ is overlap connected. Obviously $\mcO:=\{\mbO(x):x\in\llb X\rrb\}$ is an $\llb X\rrb$-cover.
Any two points in $\mbO(x)$ are overlap connected, since they are both  overlap connected with $x$.
Furthermore, if any two sets $\mbO(x)$ and $\mbO(x')$ have some point $w$ in common, then they must 
coincide, since $x \conn y$ and $x'\conn y$ imply that $x\conn x'$. 
Moreover, if $\mbO(x)$ and $\mbO(z)$ are distinct, hence disjoint, then they are overlap isolated; 
otherwise some point $v$  would be overlap connected with both $x$ and $z$ and thus lie in $\mbO(x)\cap\mbO(z)$,
which is impossible.
Thus the family $\mcO$ is an overlap  partition.

To prove that it is unique, let $\mcO'$ be any  overlap partition.
Then every set $\mbO'$ in $\mcO'$ must be contained in  $\mbO(x)$,
for each $ x\in\mbO'$.
However, $\mbO(x)$ must also be included in $\mbO'$. Otherwise there would be a point $q$ outside $\mbO'$ that is overlap connected with $x$;
this $q$ would have to lie in some set $\mbQ\in\mcO'\setminus\{\mbO\}$, impossible since $\mbQ$ must be overlap isolated from $\mbO'$.
Thus $\mbO'=\mbO(x)$ for each $x\in\mbO'$, and so $\mcO'=\{\mbO(x)\}=\mcO$. 

To establish (\ref{overlapsets}), for any  $\mbC\in\mcO$ 
let $\mbD:=\{y\in\llb Y\rrb: \llb X|y\rrb\conn\mbC\}$.
As each element of $\mcO$ consists of  all the points it is overlap connected with,
it follows that  $\llb X|y\rrb\subseteq\mbC$, for each $y\in\mbD$. 
Furthermore  $\llb X|y'\rrb$ and $\mbC$ are overlap isolated
and thus have null intersection, for every $y'\in\llb Y\rrb\setminus\mbD$. 
Thus
\bean
\mbC &=& \bigcup_{\mbB\in\llb X|Y\rrb}\mbC\cap\mbB\\
&=& \bigcup_{\mbB\in\llb X|Y\rrb:\mbB\conn\mbC}\mbC\cap\mbB 
= \bigcup_{\mbB\in\llb X|Y\rrb:\mbB\conn\mbC}\mbB.
\eean

To prove (\ref{maxcard}), observe that every set 
 $\mbC\in\llb X|Y\rrb_*$
intersects exactly one set $\mbP_\mbC\in\mcP$, i.e. $\mbP_\mbC\supseteq\mbC$.
Otherwise, $\mbC$ would also overlap some other set $\mbP'\neq\mbP_\mbC$ in the partition $\mcP$;
since $\mbC$ is overlap-connected,
this would imply that there is a point in $\mbP_\mbC$ and one in $\mbP'$ that are overlap-connected,
 which is impossible since $\mcP$ is an overlap-isolated partition.
Furthermore, since  $\llb X|Y\rrb_*$is a cover of $\llb X\rrb$, every set in $\mcP$ must intersect some set in it.  
Thus $\mbC\mapsto\mbP_\mbC$ is a surjection from $\llb X|Y\rrb_*\to\mcP$ and so  $|\llb X|Y\rrb_*|\geq |\mcP|$.

To prove the equality condition, observe that $\forall \mbP\in\mcP$, 
\[
\mbP = \bigcup_{\mbC\in\llb X|Y\rrb_*}\mbC\cap\mbP
=  \bigcup_{\mbC\in\llb X|Y\rrb_*: \mbC\cap\mbP\neq\emptyset }\mbC.
\]
If $|\llb X|Y\rrb_*|= |\mcP|$, then $\mbC\mapsto\mbP_\mbC$ is a bijection from $\llb X|Y\rrb_*\to\mcP$, 
and so the union above can only run over one set $\mbC$. 
Consequently $\mbP_\mbC=\mbC$, i.e. the bijection $\mbC\mapsto\mbP_\mbC$ from $\llb X|Y\rrb_*\to\mcP$ is
an identity. 

\section{Proof of Lemma  \ref{connectlem}
}
\label{connectlempf}

With regard to the first statement, note that
if  $(x,y),(x',y')\in\llb X,Y\rrb$ are taxicab connected,
then there is a taxicab sequence 
\[
(x,y_1),(x_2,y_1),(x_2,y_2),(x_3,y_2),\ldots, (x_{n-1},y_{n-1}), (x',y_{n-1})
\]
of points  
in $\llb X,Y\rrb$.
This yields a sequence $\{\llb X|y_i\rrb\}_{i=1}^{n-1}$ of conditional ranges 
s.t. $x_i\in \llb X|y_i\rrb\cap\llb X|y_{i-1}\rrb\neq\emptyset$ for each $ i\in [2,\ldots , n-1]$,
with $x\in \llb X|y_1\rrb$ and $x'\in \llb X|y_{n-1}\rrb$.
Thus $x\conn x'$. 

To prove the reverse implication, suppose that  $x\conn x'$ 
and pick any $y\in\llb Y|x\rrb$ and $y'\in\llb Y|x'\rrb$.
Then $\exists$ a sequence $\{\llb X|y_i\rrb\}_{i=1}^{n}$ of conditional ranges 
s.t. $\llb X|y_i\rrb\cap\llb X|y_{i-1}\rrb\neq\emptyset$, for each $i\in [2,\ldots , n]$,
where $y_1=y$ and $y_n=y'$.
For every $i\in [2,\ldots , n]$ pick an $x_i\in\llb X|y_i\rrb\cap\llb X|y_{i-1}\rrb$.
Then the taxicab sequence 
\[
(x,y_1),(x_2,y_1),(x_2,y_2),(x_3,y_2),\ldots, (x_{n},y_{n}), (x',y_{n})
\]
comprises points  
in $\llb X,Y\rrb$.
Thus  $(x,y),(x',y')$ are taxicab connected in $\llb X,Y\rrb$.

To prove the forward implication of the 2nd statement, note that  
if any $(x,y)\in\mbA$ is taxicab connected with any  $(x',y')\in\mbA$,
then $x,x'\in\mbA^+$ are overlap connected.
Similarly,  if every  $x,x'\in\mbA^+$ are overlap connected 
then for each $y\in\llb Y|x\rrb$ and $y'\in\llb Y|x'\rrb$,
$(x,y)$ is taxicab connected with  $(x',y')$.
The statement then follows by noting that $\mbA\subseteq \bigcup_{y\in\llb Y|x\rrb, x\in\mbA^+}\{(x,y)\}$.

The 3rd statement ensues similarly.
If every  $(x,y)\in\mbA$ is taxicab disconnected from any  $(x',y')\in\mbB$,
then every $x\in\mbA^+$ is overlap disconnected from any  $x'\in\mbB^+$.

Similarly,  if every  $x\in\mbA^+$ is overlap disconnected from any  $x'\in\mbB^+$,
then $\forall y\in\llb Y|x\rrb$ and $y'\in\llb Y|x'\rrb$,
$(x,y)$ is taxicab disconnected from  $(x',y')$.
The proof is completed by noting that $\mbA\subseteq\bigcup_{y\in\llb Y|x\rrb, x\in\mbA^+}\{(x,y)\}$
and  $\mbB\subseteq\bigcup_{y'\in\llb Y|x'\rrb, x'\in\mbA^+}\{(x',y')\}$.

\section{Proof of Theorem  \ref{partitionthm} (Unique Taxicab Partition)}
\label{partitionthmpf}

 For any set $\mbC$ in the unique overlap partition $\llb X|Y\rrb_*$,
define $\mbC^-:=\bigcup_{y\in\llb Y|x\rrb, x\in\mbC}\{(x,y)\}\subseteq\llb X,Y\rrb$ 
and the cover $\mcC^-:=\{\mbC^-:\mbC\in\llb X|Y\rrb_*\}$ of $\llb X,Y\rrb$. 

By Lemma \ref{connectlem},  the sets of $\mcC^-$ are individually taxicab connected
and mutually taxicab isolated, so $\mcC^-$
is a taxicab partition.

To establish uniqueness, note that if $\mcP$ is any taxicab partition, then by the same token
its projection is an overlap partition, 
which by uniqueness  must coincide with  $\llb X|Y\rrb_*$.
Thus $\forall \mbP\in\mcP$,
\[
\mbP\subseteq \bigcup_{y\in\llb Y|x\rrb, x\in\mbP^+}\{(x,y)\}
\in\mcC^-
\]
i.e. every set in $\mcP$ is inside a single set in $\mcC^-$. 
As $\mcP$ and $\mcC^-$ are partitions of  $\llb X,Y\rrb$,
it follows then that $\mbP$ must coincide exactly with an element of $\mcC^-$.

To prove (\ref{maxcardtaxi}), first observe that every set 
 $\mbD\in\mcT[X;Y]$
intersects exactly one set $\mbQ_\mbD\in\mcQ$, i.e. $\mbQ_\mbD\supseteq\mbD$.
Otherwise, $\mbD$ would also intersect some other set $\mbQ'\neq\mbQ_\mbD$ in the partition $\mcQ$;
since $\mbD$ is taxicab-connected, this would imply that there is a point in $\mbQ_\mbD$ and one in $\mbQ'$ that are taxicab-connected,
 which is impossible since $\mcQ$ is a taxicab-isolated partition.
Furthermore, since $\mcT[X;Y]$ is a cover, every set in $\mcQ$ must intersect some set in it. 
Thus $\mbD\mapsto\mbQ_\mbD$ is a surjection from $\mcT[X;Y]\to\mcQ$ and so  $|\mcT[X;Y]|\geq |\mcQ|$. 

\section{Proof of Lemma  \ref{invlem}}
\label{invlempf}

Pick any $y(0:t)\in\mbY^{t+1}$. As $\llb  X_i|y_i\rrb   \stackrel{(\ref{defR})}{\subseteq}\mbR(y_i)$ for each $i\in[0,\ldots , t]$, 
it follows that $\llb  X(0:t)|y(0:t)\rrb    \subseteq \prod_{i=0}^t \llb  X_i|y_i\rrb   $ $ \subseteq \prod_{i=0}^t\mbR(y_i)$.
Moreover  $\llb  X(0:t)|y(0:t)\rrb   \subseteq\llb  X(0:t)\rrb   $, thus establishing that the LHS of (\ref{invchannel}) is contained in the RHS.

It is now shown that the RHS is contained in the LHS,  proving equality.
If the RHS is empty then so is the LHS, by the preceding argument,  yielding the desired equality.
If the RHS is not empty,  pick an arbitrary element $x(0:t)$ in it,
i.e. $x(0:t)\in\llb  X(0:t)\rrb   $ and  $x(i)\in\mbR(y(i))$, for each $ i\in[0,\ldots ,t]$.
By (\ref{defR}), $y(i)\in\mbT(x(i))$ for each $ i\in[0,\ldots ,t]$, or equivalently 
$y(0:t)\in\prod_{i=0}^t\mbT(x_i)$ $\stackrel{(\ref{defchannel})}{=} \llb  Y(0:t)|x(0:t)\rrb   $. 
Thus $\exists\ome\in\Ome$ s.t. $Y(0:t)(\ome)=y(0:t)$ and $X(0:t)(\ome)=x(0:t)$.
This implies that $x(0:t)\in\llb  X(0:t)|y(0:t)\rrb   $. Thus the RHS of (\ref{invchannel}) is contained in the LHS,
completing the proof.

%

\end{document}